\documentclass[a4paper, pra, twocolumn, superscriptaddress]{revtex4-1}
\usepackage{booktabs}
\usepackage{physics}
\usepackage{graphicx}
\usepackage{lipsum}
\usepackage{tikz}
\usepackage{quantikz}
\usepackage{setspace}   
\usepackage{hyperref} 
\usepackage{amsmath, amssymb}
\usepackage[normalem]{ulem}

\newcommand{\be}{\begin{equation}}
\newcommand{\ee}{\end{equation}}
\newcommand{\bea}{\begin{eqnarray}}
\newcommand{\eea}{\end{eqnarray}}

\begin{document}

\newcommand{\uv}{Departament de Física Teòrica and IFIC, Universitat de València-CSIC, 46100 Burjassot (València), Spain}

\newcommand{\ific}{IFIC, Universitat de València – CSIC, Parque Científico, Catedrático José Beltrán 2, 46980 Paterna (València), Spain}

\newcommand{\ift}{Instituto de Física Teórica, IFT-UAM/CSIC\\
Universidad Autónoma de Madrid, Cantoblanco, 28049 Madrid, Spain}

\title{Variational noise mitigation in quantum circuits: \\the case of Quantum Fourier Transform}

\author{Rafael Gómez-Lurbe}
\affiliation{\uv}
\author{Alexander Bernal}
\affiliation{\ift}
\author{Armando Pérez}
\affiliation{\uv}
\author{Bryan Zaldívar}
\affiliation{\ific}
\author{J. Alberto Casas}
\affiliation{\ift}

\begin{abstract}

We propose using variational quantum algorithms (VQAs) to simulate established quantum algorithms under realistic noise conditions, aiming to surpass the fidelity of theoretical circuits in noisy environments. Focusing on the Quantum Fourier Transform (QFT), we perform numerical simulations for two qubits under both coherent and incoherent noise. To enhance generalization, we further introduce the use of Mutually Unbiased Bases (MUBs) during the optimization. Our results show that the variational circuit can reproduce the QFT with higher fidelity in scenarios dominated by coherent noise. This demonstrates the potential of the approach as an effective error-mitigation strategy for small- to medium-scale quantum systems, particularly in settings where coherent noise strongly impacts performance. Beyond mitigating noise and improving fidelity, the method can be adapted to the noise profile of a specific device, providing a versatile and practical route to enhance the reliability of quantum algorithms in near-term quantum hardware.

\end{abstract}
\maketitle

\section{Introduction}\label{Introduction}
Quantum computing represents a paradigm shift in computational science, enabling the solution of problems that are unfeasible for classical algorithms by exploiting principles of quantum mechanics such as superposition, entanglement, and quantum interference~\cite{nielsen2010quantum}. Among the most prominent quantum algorithms offering a theoretical advantage over their classical counterparts is the Quantum Fourier Transform (QFT)~\cite{coppersmith1994qft}. QFT plays a crucial role in numerous quantum algorithms. Notably, it is a core component of Shor's algorithm for integer factorization and computing discrete logarithms \cite{shor1997polynomial}. It is also integral to the quantum phase estimation algorithm, which is used for estimating the eigenvalues of unitary operators \cite{nielsen2010quantum}. Additionally, the QFT is central to algorithms designed to solve instances of the Hidden Subgroup Problem, a class of problems that encompasses many computational challenges in cryptography and number theory \cite{lomont2004hidden}.

Despite these promising theoretical results, practical implementations of such algorithms on current quantum devices remain a significant challenge due to hardware limitations and the prevalence of errors. Presently, we are in the so-called Noisy Intermediate-Scale Quantum (NISQ) era~\cite{preskill2018nisq}, characterized by quantum processors with a moderate number of qubits (on the order of hundreds) that are susceptible to various noise sources. These limitations severely restrict the depth and accuracy of quantum computations, making it difficult to achieve the fault-tolerance required for implementing complex quantum algorithms like Shor's.

In this context, Variational Quantum Algorithms (VQAs) have emerged as a promising approach to harness the computational capabilities of NISQ devices~\cite{cerezo2021variational}. VQAs are hybrid quantum-classical algorithms in which a quantum computer is employed to prepare parameterized quantum states, and the parameters are optimized using classical optimization routines based on measurements performed on the quantum hardware. This hybrid approach allows VQAs to mitigate some of the limitations of current quantum devices, making them a practical tool for exploring quantum advantage in tasks such as optimization, machine learning, and quantum simulation.

In this work, we aim to utilize variational quantum circuits to simulate established quantum circuits under various noise conditions, with the objective of achieving higher fidelity than the original circuits in the presence of noise. Specifically, we focus on the Quantum Fourier Transform and perform numerical simulations for a 2-qubit system across different noise scenarios, including both coherent and incoherent noise sources. To improve the generalization of the trained model, we propose the use of Mutually Unbiased Bases (MUBs) during the optimization, rather than restricting the training solely to computational basis states. This strategy exploits the structure of quantum state space more comprehensively, allowing the variational circuit to capture a broader range of input states and thereby achieve improved generalization across a wider class of inputs. As a result, the circuit is better equipped to reproduce the QFT accurately under different noise conditions.



Our primary objective is to develop a protocol tailored to the specific noise profile of a given quantum device, enabling the design of a variational quantum circuit that reproduces the QFT with fidelity higher than that of the original quantum circuit under realistic noisy conditions. Such a protocol would serve as a valuable tool for certifying that a quantum device is capable of preparing the QFT with enhanced fidelity. Importantly, the proposed protocol requires access to both the classically computed ideal QFT state vectors and the noisy output states—represented by the density matrix by the variational circuit. These elements are compared within a classical optimization loop to train the circuit parameters using a realistic noise model. Consequently, the approach serves as a practical certification and noise-mitigation tool for small- to medium-scale quantum systems, where full classical simulation of the QFT remains feasible.

Furthermore, the protocol is particularly effective in regimes dominated by coherent noise, where variational adaptation can significantly enhance fidelity. Beyond its use for device certification, the resulting variationally optimized circuits can be embedded as subroutines within larger quantum algorithms, thereby improving their robustness to hardware-specific noise. An explicit case is for instance the multiqubit QFT, which can be recursively constructed from the 2-qubit case \cite{Cleve2000}.

The work is organized as follows. In Sec.~\ref{Theory}, we briefly review the theoretical background, including the variational quantum circuit employed for the simulations, the optimization procedure, and the noise models considered for the numerical experiments. Subsequently, Sec.~\ref{Results} presents the numerical experiments conducted under the different noise scenarios. Finally, in Sec.~\ref{Conclusions}, we summarize the conclusions drawn from this work and outline future directions and open problems in this research area.

\section{Theory}\label{Theory}

As mentioned above, the aim of this project is to achieve a higher fidelity for the QFT in the presence of noise by using a variational circuit, as a replacement of the QFT itself.

In this section, we first present the 2-qubit QFT and the corresponding variational quantum circuit used to simulate it. Subsequently, we describe the classical optimization setup for tuning the parameters of the variational circuit, detailing the optimizer and cost function employed. Lastly, we outline the different noise scenarios considered in our numerical simulations.

The 2-qubit investigation can be considered, not only as a case study but also as a first step for the analysis of the QFT with a larger number of qubits. In fact, as discussed in \cite{Cleve2000}, the QFT circuit can be built recursively, where the setup $\mathcal{F}_{2^{n}}$ for $n$ qubits is obtained by first applying $\mathcal{F}_{2^{n-1}}$ to the first $n-1$ qubits, followed by the action of the phase gate $P(1/2^{n-j+1})$ on the $j^{th}$ and $n^{th}$ qubits, with $j\in\{1,2,\dots n-1\}$, and a Hadamard $H$ gate applied to the $n^{th}$ qubit.

\subsection{2-Qubit Quantum Fourier Transform and variational model}\label{QFT}

\begin{figure}
        \centering
  \text{Quantum circuit for 2-Qubit QFT}  \\ \vspace{0.2cm}
        \includegraphics[scale=0.35]{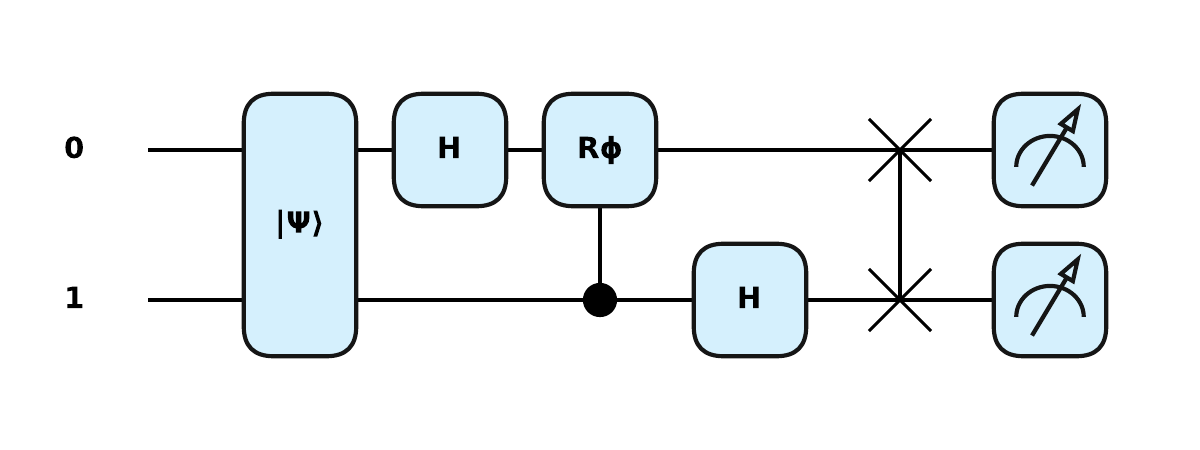}

         \caption{The quantum circuit for the 2-qubit QFT, where \(|\psi\rangle\) denotes the initial state before applying the QFT, implemented using the Pennylane library.}
    \label{fig:QFT}
  
\end{figure}

We focused our study on the 2-qubit Quantum Fourier Transform. For the circuit implementation and simulation, we employed the \texttt{Pennylane} library \cite{bergholm2022pennylaneautomaticdifferentiationhybrid}, which enables efficient construction, simulation, and optimization of quantum circuits. The circuit implementation is depicted in Fig.~\ref{fig:QFT}.

The 2-qubit QFT consists of a sequence of gates that include the Hadamard gate, the SWAP gate, and the controlled-phase (\( \text{CPhase} \)) gate. The controlled-phase gate is defined as
\begin{equation}
\text{CPhase}(\phi) = 
\begin{bmatrix}
1 & 0 & 0 & 0 \\
0 & 1 & 0 & 0 \\
0 & 0 & 1 & 0 \\
0 & 0 & 0 & e^{i\phi}
\end{bmatrix},
\end{equation}
where \( \phi \) is the phase angle applied to the controlled state.

The input state for the QFT is denoted as \( |\psi\rangle \), which may represent, for example, a computational basis state.

To simulate the QFT, we propose an Ansatz for the variational quantum circuit inspired by the decomposition of an arbitrary 2-qubit controlled gate \cite{benenti2004principles}:
\[
\begin{quantikz}[scale=0.1]
\lstick{} & \ctrl{1}  & \rstick{} \\
\lstick{} & \gate{U}  & \rstick{}
\end{quantikz}
\quad = \quad
\begin{quantikz}[scale=0.1]
\lstick{} & \qw & \ctrl{1} & \qw & \ctrl{1}  & \gate{\delta} & \qw \\
\lstick{} & \gate{C} & \targ{} & \gate{B} & \targ{} & \gate{A} & \qw 
\end{quantikz},
\]

where \( U \) is a general single-qubit unitary defined as
\begin{equation}
U =
\begin{bmatrix}
e^{i(\delta - \alpha/2 - \beta/2)} \cos\!\frac{\theta}{2} &
- e^{i(\delta - \alpha/2 + \beta/2)} \sin\!\frac{\theta}{2} \\[4pt]
e^{i(\delta + \alpha/2 - \beta/2)} \sin\!\frac{\theta}{2} &
e^{i(\delta + \alpha/2 + \beta/2)} \cos\!\frac{\theta}{2}
\end{bmatrix},
\end{equation}
and the single-qubit gates are given by
\begin{align}
A &= R_z(\alpha)\, R_y\!\left(\frac{\theta}{2}\right), \\
B &= R_y\!\left(-\frac{\theta}{2}\right)\, R_z\!\left(-\frac{\alpha+\beta}{2}\right), \\
C &= R_z\!\left(\frac{\beta-\alpha}{2}\right), \\
\delta &= R_z(\delta)
\end{align}
Here, \(\alpha\), \(\beta\), \(\theta\), and \(\delta\) are real parameters that fully determine the unitary transformation, while \(R_x\), \(R_y\), and \(R_z\) denote the standard single-qubit Pauli rotation operators.

\begin{figure}
        \centering
  \text{Variational Ansatz}  \\ \vspace{0.2cm}
        \includegraphics[scale=0.35]{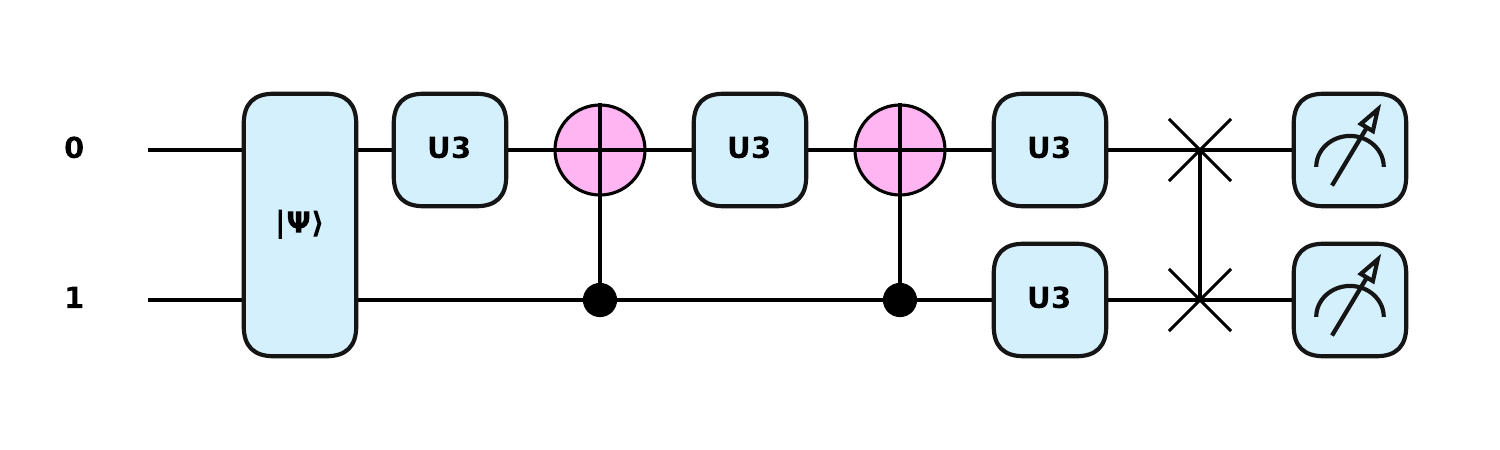}

         \caption{The quantum circuit for the 2-qubit variational Ansatz used to simulate the QFT, where \(|\psi\rangle\) denotes the initial state before applying the Ansatz, implemented using the Pennylane library.}
    \label{fig:Variational}
  
\end{figure}

The quantum circuit implementation of the Ansatz in \texttt{Pennylane} is shown in Fig.~\ref{fig:Variational}.

Namely, the variational circuit consists of a sequence of arbitrary single-qubit unitary gates \( U_3 \), CNOT gates, and SWAP gates. The \( U_3 \) gate is defined as:
\begin{equation}
U_3(\theta, \phi, \lambda) = 
\begin{bmatrix}
\cos\left(\frac{\theta}{2}\right) & -e^{i\lambda} \sin\left(\frac{\theta}{2}\right) \\
e^{i\phi} \sin\left(\frac{\theta}{2}\right) & e^{i(\phi + \lambda)} \cos\left(\frac{\theta}{2}\right)
\end{bmatrix},
\end{equation}
where \( \theta \), \( \phi \), and \( \lambda \) are real parameters that define the rotation angles of the qubit.
With this Ansatz, we have a set of 12 trainable parameters corresponding to the three parameters of each of the four $U_3$ gates used in the circuit.

\subsection{Classical Optimization}\label{Optimization}
The optimization of the trainable parameters is performed classically using the gradient descent method. Gradient descent is a standard optimization algorithm used to minimize a function by iteratively moving in the direction of the steepest descent, defined as the negative gradient of some cost function. It is widely used in machine learning and statistics to optimize models by adjusting their parameters \cite{ruder2016overview}.

To define the cost function, we first recall that the objective is to design a variational circuit that approximates the ideal quantum Fourier transform ($QFT^{\text{ideal}}$) for any initial state.
We can distinguish two cases. 

The first one is the noiseless case, where we consider that the variational circuit is not affected by any kind of noise. In this idealized scenario, it is important to note that the training process requires access to the state vectors produced by both the ideal QFT and the noiseless variational circuit. These are computed classically and compared during optimization to define the cost function. Naturally, this is an ideal scenario, which does not occur in practice, but it is useful, as we will see, to verify that, in that limit, the variational circuit is capable of emulating the ideal QFT operation. 
In this case, the action of the circuit is described by some unitary operation, $U(\boldsymbol{\theta})$, where $\boldsymbol{\theta}$ collectively represents the tunable parameters of the variational circuit. Hence, training the circuit to perform the QFT on a set of basis states is sufficient since any other pure initial state is given by a superposition of these elements. For simplicity, we have taken the computational basis $\{\ket{i}\}_{i=1}^4$. Then, the cost function is defined as: 
\begin{equation}\label{eq:costPure}
C(\boldsymbol{\theta}) = \frac{1}{4} \sum_{i=1}^4 \left\| U(\boldsymbol{\theta})\ket{i} - QFT^{\text{ideal}}\ket{i} \right\|^2.
\end{equation}
where the distance between the states is measured by the Euclidean norm. We train with this cost function rather than fidelity, since the latter is insensitive to global phases. Training directly with fidelity would allow the circuit to learn each basis state only up to an arbitrary phase, potentially assigning different phases to different basis states. As a result, when applied to a superposition, the circuit could yield an incorrect state differing from the true QFT output.

Nevertheless, the fidelity still gives a powerful tool to assess the quality of the state produced by the variational circuit. Hence, alongside the cost function we also compute its fidelity with respect to the corresponding state from the ideal QFT. In this noiseless case, the average fidelity reads
\begin{equation}\label{eq:fidPure}
F_{avg.}(\boldsymbol{\theta}) = \frac{1}{4} \sum_{i=1}^4 \abs{\mel{i}{U^\dagger(\boldsymbol{\theta}) QFT^{\text{ideal}}}{i}}^2.
\end{equation}

\vspace{0.2cm}
The second case corresponds to the presence of noise in the variational circuit and is therefore a much more realistic scenario. Then the map  implemented by the circuit is in general non-unitary, leading to a mixed state described by a density matrix. Hence, learning just from the basis states as input is no longer sufficient to correctly implement the QFT on an arbitrary state. The variational circuit would converge to a set of parameters such that 
\begin{equation}
\mathcal{N}^{\text{var}}_{\boldsymbol{\theta}}(|i\rangle\langle i|) \simeq QFT^{\text{ideal}}(|i\rangle\langle i|)(QFT^{\text{ideal}})^{\dagger}.
\end{equation}
where $\mathcal{N}^{\text{var}}_{\boldsymbol{\theta}}$ is the circuit map.
However, this equivalence does not extend in general to off-diagonal elements of the density matrix:
\begin{equation}
    \mathcal{N}^{\text{var}}_{\boldsymbol{\theta}}(|i\rangle\langle j|) \not\simeq QFT^{\text{ideal}}(|i\rangle\langle j|)(QFT^{\text{ideal}})^{\dagger}, \quad \text{for } i \neq j.
\end{equation}

Of course, training the variational circuit on every possible input state is practically infeasible due to the continuous and infinite nature of the state space. Instead, we train it using a representative subset of states from the Hilbert space. 

In this sense, a natural and effective alternative is to train the variational circuit using \textit{Mutually Unbiased Bases} (MUBs) \cite{Durt:2010egm}. This strategy exploits the structure of quantum state space more comprehensively and achieves improved generalization across a wider class of input states.

We briefly remind the notion of MUBs. Namely, two bases are called mutually unbiased if the pairwise overlap between elements from each basis is constant. In the same fashion, a set of $m$ bases $\{M_x\}_{x=1}^m$, each of them conformed by the basis elements $M_x=\{\ket*{e_a^{(x)}}\}_{a=1}^d$, are MUBs if $\left|\langle e^{(x)}_a|e^{(x')}_{a'}\rangle \right|^2={1}/{d}$ for $x\neq x'$ and for any $a,a'$. Given any dimension $d$, at least $m=3$ and at most $m=d+1$ MUBs exist. In particular, for two qubits ($d=4$) the upper bound is saturated by $m=5$, and a set of 5 MUBs reads 
\begin{equation}
\begin{aligned}
     M_1=&\{(1,0,0,0),(0,1,0,0), (0,0,1,0), \\
                         &(0,0,0,1)\}, \\
     M_2=&\frac{1}{2}\{(1,1,1,1), (1,1,-1,-1),(1,-1,-1,1),\\
                         &(1,-1,1,-1)\}, \\
     M_3=&\frac{1}{2}\{(1,-1,-i,-i),(1,-1,i,i), (1,1,i,-i),\\                      &(1,1,,-i,i)\}, \\
     M_4=&\frac{1}{2}\{(1,-i,-i,-1),(1,-i,i,1), (1,i,i,-1),\\                      &(1,i,-i,1)\}, \\
     M_5=&\frac{1}{2}\{(1,-i,-1,-i),(1,-i,1,i),(1,i,-1,i), \\                      &(1,i,1,-i)\}.
\end{aligned}\label{eq:MUBs}
\end{equation}

Now we define the cost function for the noisy case as:
\begin{equation}
  C(\boldsymbol{\theta}) = \frac{1}{20} \sum_{a,x \in \text{MUBs}} \left\| \rho_a^{(x)}(\boldsymbol{\theta}) - q_{a}^{(x)}\right\|^2,  
\end{equation}
where 
\bea
\rho_a^{(x)}(\boldsymbol{\theta})&\equiv&\mathcal{N}^{\text{var}}_{\boldsymbol{\theta}}(\ketbra*{e^{(x)}_a})
\nonumber\\
q_{a}^{(x)}&\equiv& QFT^{\text{ideal}}(\ketbra*{e^{(x)}_a})(QFT^{\text{ideal}})^{\dagger},
\eea
and distance between the matrices is measured using the Frobenius norm: $\|M\|=\sqrt{\tr(M^\dagger M)}$. 

In this noisy scenario, the optimization requires access to the density matrix that represents the output of the variational circuit under noise, as well as the corresponding density matrix obtained from the ideal QFT. Both are evaluated classically within the cost function to guide the parameter updates under the chosen noise model.

Similarly to the noiseless case we also compute the average fidelity, which for mixed states is defined by:
\begin{equation}
  F_{avg}(\boldsymbol{\theta}) = \frac{1}{20} \sum_{a,x \in \text{MUBs}} \left\| \sqrt{\rho_a^{(x)}(\boldsymbol{\theta}) } \sqrt{q_{a}^{(x)}}\right\|_{\text{tr}}^2,  
\end{equation}
where $\left\|M\right\|_{\text{tr}}=\tr(\sqrt{M^\dagger M})$.

\subsection{Noise scenarios}\label{Noise_Scenarios}

The ultimate goal of our study is to analyze the ability of the variational circuit to reproduce the QFT in the presence of different sources of noise, with the aim of achieving a better fidelity than the original QFT circuit. 

In this work, we consider four different noise scenarios:
\begin{enumerate}
    \item Noiseless scenario.
    \item Depolarizing noise.
    \item Depolarizing noise with single-qubit thermal relaxation.
    \item Depolarizing noise with crosstalk between qubits.
    \item Depolarizing noise with crosstalk between qubits and single-qubit thermal relaxation.
\end{enumerate}

\subsubsection{Noiseless scenario}\label{Noiseless_theo}

The assumption of the absence of noise is, of course, unrealistic, but, as mentioned above, we have considered this scenario in the first place as a check that our noiseless variational circuit is capable of reproducing the standard QFT.

\subsubsection{Depolarizing noise}\label{DP_theo}

Depolarizing noise is one of the most commonly used noise channels to study the effects of incoherent (non-unitary) noise. The \( n \)-qubit depolarizing channel is defined as:

\begin{equation}
\mathcal{D}_n[\epsilon](\rho) = (1 - \epsilon)\rho + \frac{\epsilon}{2^n}I,
\end{equation}
where \( \rho \) is the quantum state prior to the effect of noise, \( \epsilon \) is the depolarization probability, and \( I \) is the \( n \)-qubit identity matrix.

In our simulations, we applied a single-qubit depolarizing channel with parameter \( \epsilon = 2.5 \cdot 10^{-4} \) after each single-qubit gate, and a two-qubit depolarizing channel with parameter \( \epsilon = 2.5 \cdot 10^{-3} \) after each two-qubit gate, i.e., ten times the single-qubit value. These parameters were chosen based on calibration data from state-of-the-art quantum devices \cite{ibm_quantum_2025}.

Although symmetric depolarizing noise is uncommon in real devices, this model provides a useful framework for exploring very general incoherent noise effects.

\subsubsection{Depolarizing noise with single-qubit thermal relaxation}

To accurately reflect the noise characteristics of real quantum hardware, we adopt a composite noise model consisting of a depolarizing channel followed by a single-qubit thermal relaxation channel. The parameters for this model are derived from calibration data of actual quantum devices. This approach captures both energy relaxation and pure dephasing—the two primary decoherence mechanisms in superconducting qubits, such as those used in IBM Quantum systems. Notably, this is also the methodology used by IBM to model the realistic noise behavior of their quantum devices.

The single-qubit thermal relaxation noise channel is characterized by four physical parameters: the energy relaxation time $T_1$, the total dephasing time $T_2$ (which satisfies $T_2 \leq 2 T_1$), the gate time for relaxation error $t_g$, and the excited state population at thermal equilibrium $p_e$.

The excited state population \( p_e \) is estimated using a Maxwell–Boltzmann distribution, assuming a two-level system and neglecting contributions from higher excited states. It is calculated based on the qubit frequency \( f \) (in Hz) and the device operating temperature \( T \) (in millikelvin) as follows (see~\cite{PhysRevLett.114.240501}):
\begin{equation}
p_e = \frac{1}{1 + e^{\frac{h f}{k_B T \cdot 10^{-3}}}},
\end{equation}
where \( h \) is the Planck constant and \( k_B \) is the Boltzmann constant.

Using $T_1$, $T_2$, $t_g$, and $p_e$, we define the following quantities:
\begin{equation}
\begin{aligned}
e_{T_1} &= e^{-\frac{t_g}{T_1}}, \quad
e_{T_2} = e^{-\frac{t_g}{T_2}}, \\
p_{\mathrm{reset}} &= 1 - e_{T_1}, \quad
p_z = \frac{(1 - p_{\mathrm{reset}})(1 - \frac{e_{T_2}}{e_{T_1}})}{2}, \\
p_{r0} &= (1 - p_e) \cdot p_{\mathrm{reset}}, \quad
p_{r1} = p_e \cdot p_{\mathrm{reset}}, \\
p_{\mathrm{id}} &= 1 - p_z - p_{r0} - p_{r1}\ .
\end{aligned}
\end{equation}

These quantities define the quantum channel via the following Kraus operators, as used in PennyLane:
\begin{equation}
\begin{aligned}
K_0 &= \sqrt{p_{\mathrm{id}}} \, I, 
&\quad
K_1 &= \sqrt{p_z} \, Z, \\
K_2 &= \sqrt{p_{r0}} \begin{bmatrix} 1 & 0 \\ 0 & 0 \end{bmatrix}, 
&\quad
K_3 &= \sqrt{p_{r0}} \begin{bmatrix} 0 & 1 \\ 0 & 0 \end{bmatrix}, \\
K_4 &= \sqrt{p_{r1}} \begin{bmatrix} 0 & 0 \\ 1 & 0 \end{bmatrix}, 
&\quad
K_5 &= \sqrt{p_{r1}} \begin{bmatrix} 0 & 0 \\ 0 & 1 \end{bmatrix}.
\end{aligned}
\end{equation}

As previously mentioned, this noise channel captures the combined effects of amplitude damping (relaxation), thermal excitation, and dephasing arising from environmental interactions. In our simulations, the channel is implemented using \texttt{qml.ThermalRelaxationError} in PennyLane, with physical parameters obtained from real calibration data of the IBM quantum device \texttt{ibm\_brisbane} (see the table in the Appendix \ref{Calibration Data}).

Following IBM's guidelines for noise modeling based on device backend properties and calibration data, we apply a single-qubit depolarizing channel followed by a thermal relaxation channel after each single-qubit gate. For two-qubit gates, we apply a two-qubit depolarizing channel followed by independent thermal relaxation channels on each qubit.

\subsubsection{Depolarizing noise with crosstalk between qubits}\label{DP_CT_theo}

In this scenario, we applied depolarizing noise alongside crosstalk noise between qubits. 

Crosstalk noise arises from unwanted interactions between qubits during the application of a two-qubit gate. This type of noise poses a significant challenge to the execution of two-qubit operations and may limit the scalability of algorithms on near-term quantum hardware \cite{Sarovar2020detectingcrosstalk}. Since crosstalk noise can be represented by a unitary transformation, it is classified as a form of coherent noise. The interaction between qubits due to crosstalk can be modeled by the unitary gate \cite{PhysRevApplied.12.054023}:
\begin{equation}
    U_{ZZ}[\zeta] = \exp\left( -i 2\pi \zeta T \lvert 11 \rangle \langle 11 \rvert \right),
\end{equation}
where \( T \) is the duration of the two-qubit gate, and \( \zeta \) describes the crosstalk interaction strength.

Crosstalk noise was applied after each two-qubit gate in the circuit, using \( T = 660\,\text{ns} \), corresponding to a typical duration for two-qubit entangling gates in IBM quantum devices (see the table in Appendix~\ref{Calibration Data}), and \( \zeta = 1.5 \cdot 10^5\, \text{Hz} \). The value of \( \zeta \) was chosen based on \cite{9651438}, where we have adopted one of the more conservative estimates for the interaction strength. Both parameter values are consistent with those observed in real quantum hardware. Following the crosstalk operation, we applied a two-qubit depolarizing channel with \( \epsilon = 2.5 \cdot 10^{-3} \). Additionally, we applied a single-qubit depolarizing channel with \( \epsilon = 2.5 \cdot 10^{-4} \) after each single-qubit gate.

\subsubsection{Depolarizing noise with crosstalk between qubits and single-qubit thermal relaxation}

To thoroughly assess the performance of the variational algorithm in enhancing the theoretical QFT circuit, we conducted simulations that included all relevant noise sources: depolarizing noise with crosstalk between qubits, as well as single-qubit thermal relaxation.

The noise parameters used in this setting are the same as those from the previous section, derived from real calibration data of the IBM quantum device \texttt{ibm\_brisbane} (see the table in Appendix~\ref{Calibration Data}).

Specifically, after each single-qubit gate, we apply a single-qubit depolarizing channel followed by a single-qubit thermal relaxation channel. For two-qubit gates, crosstalk noise is first applied, followed by a two-qubit depolarizing channel, and then by independent single-qubit thermal relaxation channels on each qubit.

\section{Numerical results}\label{Results}

In this section, we present the results of the numerical experiments conducted under the five different noise scenarios described in Sec.~\ref{Noise_Scenarios}. For each experiment, we plot the evolution of the cost function during optimization, along with the average fidelity relative to the ideal state produced by the QFT. These averages are computed over the MUB states.

As a validation step to confirm that our variational circuits accurately implement the QFT, we also compute the average fidelity over 1000 randomly chosen initial superposition states after optimization. To identify potential outliers, we include the full distribution of fidelity values. Additionally, we report the standard deviation of the average fidelity for all experiments.

\begin{figure}[h]
        \centering
  \text{Variational Ansatz random superposition}  \\ \vspace{0.2cm}
        \includegraphics[scale=0.32]{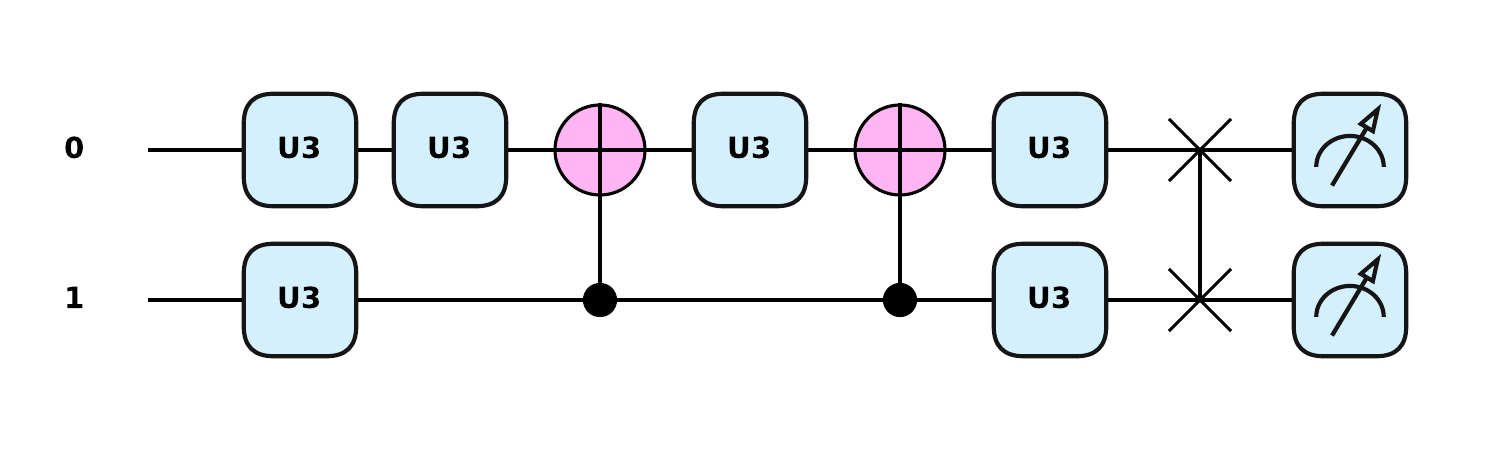}

         \caption{The quantum circuit for the variational Ansatz used to simulate the QFT with 2 qubits, with random superposition states as input, implemented using the Pennylane library.}
    \label{fig:Variational_superpo}
  
\end{figure}

The initial random superposition states are generated by applying a \( U_3 \) gate on each of the qubits and choosing the parameters randomly between 0 and \( 2\pi \) with uniform probability. The variational circuit in this setting is depicted in Fig.~\ref{fig:Variational_superpo}.

\subsection{Noiseless scenario}\label{Noiseless_Results}

\begin{figure}[]
    \centering
    \text{Cost -- Noiseless} \\ \vspace{0.2cm}
        \includegraphics[scale=0.34]{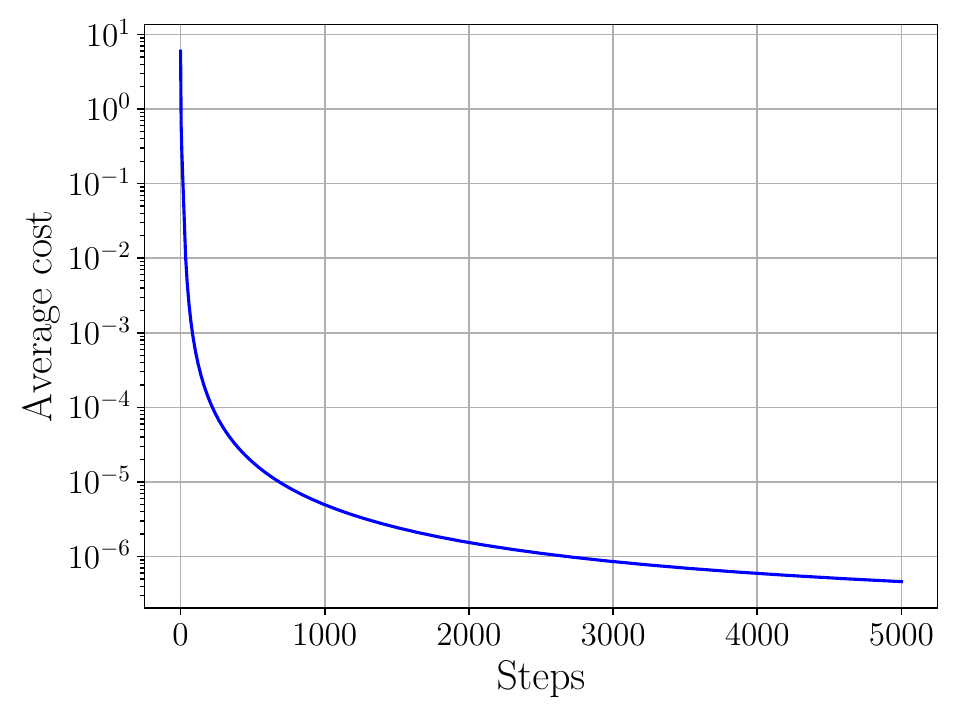}
  \caption{\small Evolution of the cost function, corresponding to the noiseless case, presented on a logarithmic scale, during optimization over 5000 steps, showing quick convergence. However, we trained for additional steps to achieve higher fidelity.}
 \label{fig:cost_noiseless}
\end{figure}

\begin{figure}[]
    \centering
    \text{Average Fidelity -- Noiseless} \\ \vspace{0.2cm}
        \includegraphics[scale=0.34]{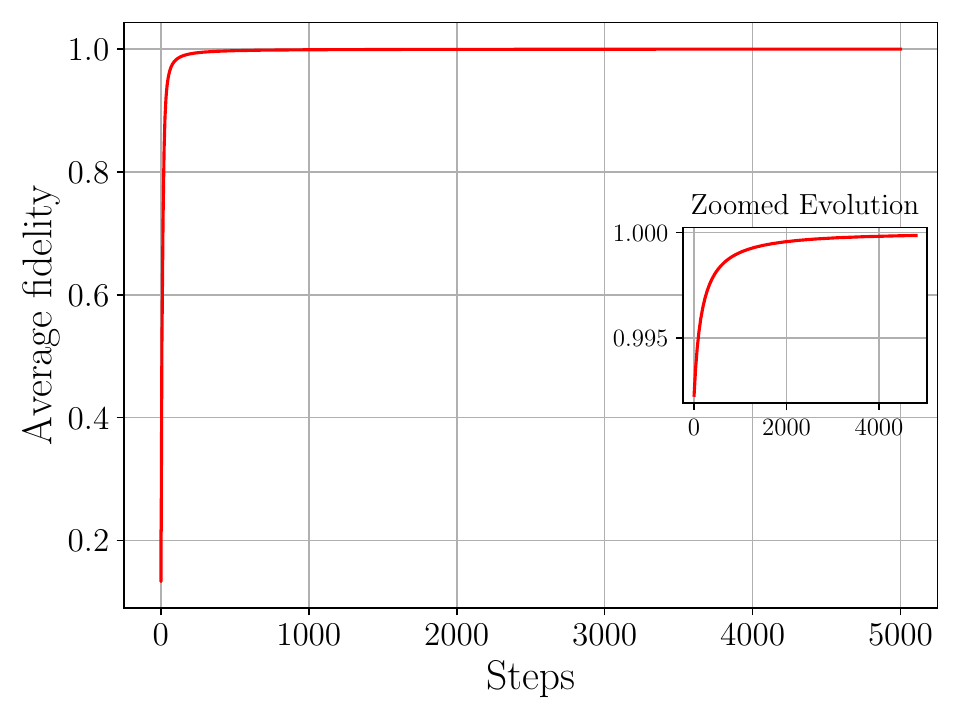}
  \caption{\small Evolution of the average fidelity during optimization. The big
  panel shows the complete optimization for 5000 steps, demonstrating quick convergence. Nevertheless, we trained for additional steps to achieve higher fidelity. The small panel displays the evolution of the fidelity after 200 steps, zooming-in.}

 \label{fig:fid_noiseless}
\end{figure}

In the noiseless setting, we performed the optimization using gradient descent for 5000 iterations with a learning rate \(\eta = 0.3\), training on the four computational basis states. The optimization was terminated around 5000 iterations when the convergence criterion, defined as \(\delta = \| C(\boldsymbol{\theta}_{t+1}) - C(\boldsymbol{\theta}_t) \| \), fell below \(10^{-9}\), indicating that the cost function had effectively converged. The evolution of the cost function, Eq.~\eqref{eq:costPure}, is shown in Fig.~\ref{fig:cost_noiseless}, while the corresponding evolution of the average fidelity, Eq.~\eqref{eq:fidPure}, is presented in Fig.~\ref{fig:fid_noiseless}. After training, the variational circuit achieves an average fidelity over the basis states of \(0.9992 \pm 0.0003\).

\begin{figure}[]
    \centering
    \text{Histogram Fidelities -- Noiseless } \\ \vspace{0.2cm}
        \includegraphics[scale=0.4]{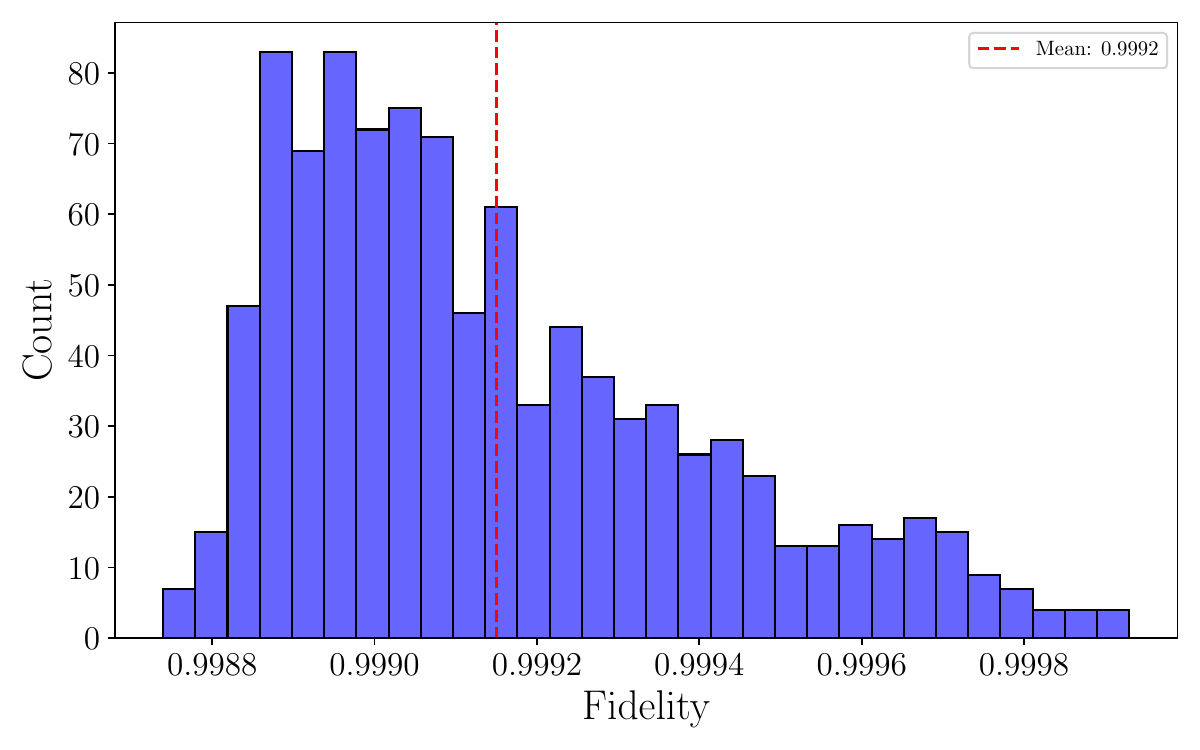}
  \caption{\small Distribution of fidelity values for 1000 random initial superposition states using the optimal parameters obtained from the optimization. The dashed red line represents the average fidelity. }

 \label{fig:distr_noiseless}
\end{figure}

After optimization, we computed the average fidelity for 1000 random initial superposition states, obtaining a fidelity of \(0.9989\) with a standard deviation of \(0.003\). Furthermore, we also plot the distribution of fidelities in Fig. \ref{fig:distr_noiseless}. This confirms that the variational circuit successfully performs the QFT with high fidelity, with no outliers present.

Notably, the matrix obtained from the circuit with the optimal parameters closely approximates the QFT matrix. Computing the Frobenius norm of the difference between these matrices results in a value of \( 0.052 \). This confirms that our variational circuit effectively implements the QFT.

\subsection{Depolarizing noise}\label{DP_results}

\begin{figure}[]
    \centering
    \text{Cost -- Depolarizing} \\ \vspace{0.2cm}
        \includegraphics[scale=0.34]{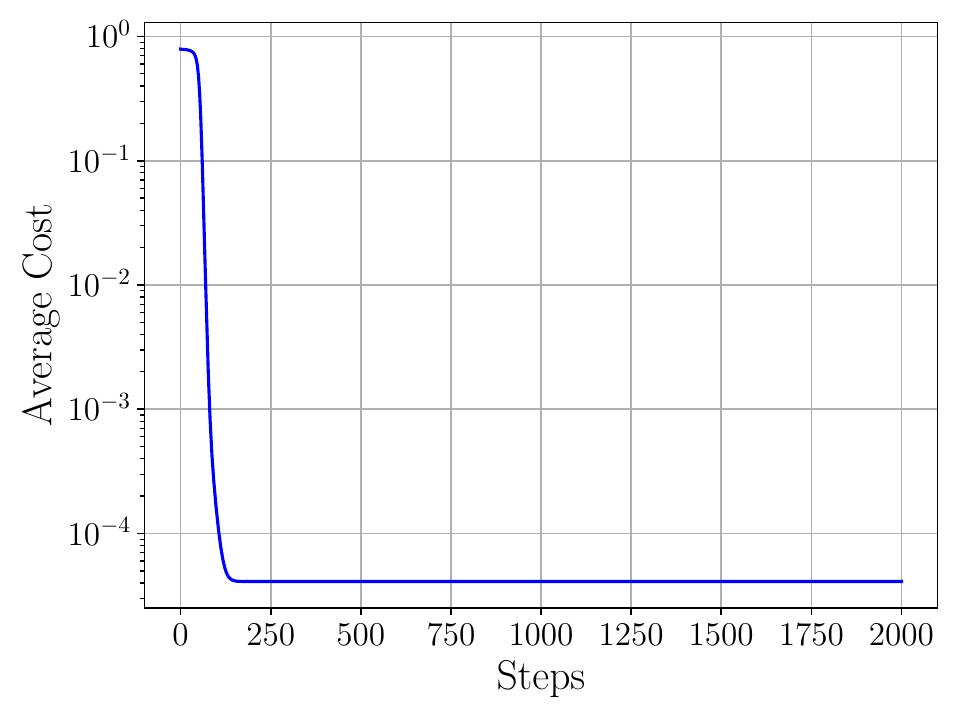}
  \caption{\small Evolution of the cost function in the presence of depolarizing noise, presented on a logarithmic scale, during optimization over 2000 steps. In this setting, training for additional steps does not achieve a higher fidelity; instead, it reaches a plateau and starts oscillating around a given value.}

 \label{fig:cost_depol}
\end{figure}

\begin{figure}[]
    \centering
    \text{Average Fidelity -- Depolarizing} \\ \vspace{0.2cm}
        \includegraphics[scale=0.34]{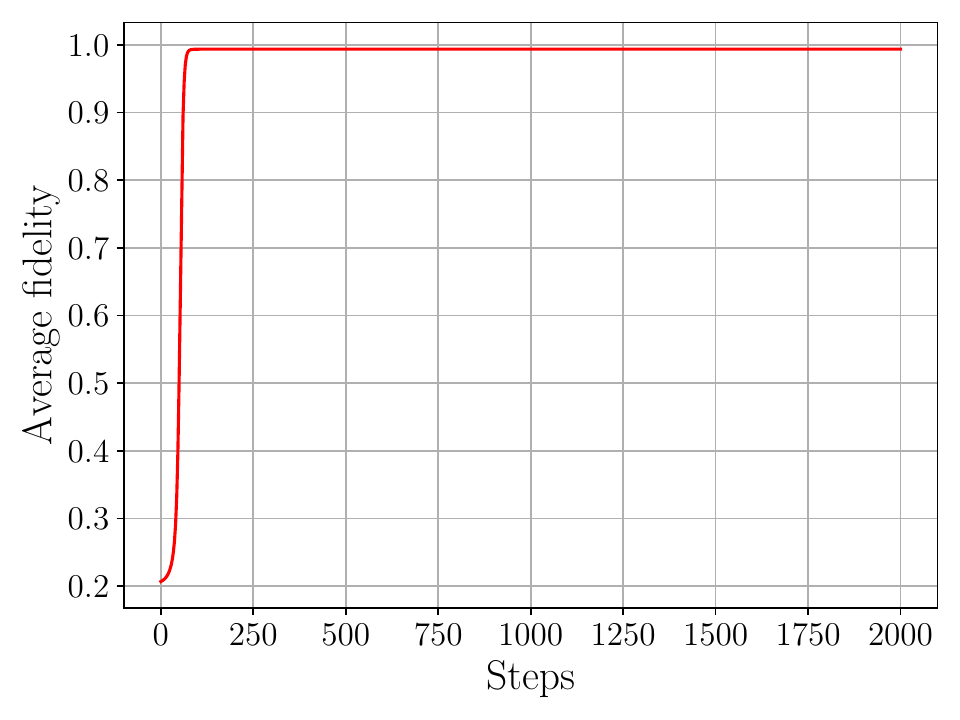}
  \caption{\small Evolution of the average fidelity during optimization in the presence of depolarizing noise over 2000 steps. In this setting, training for additional steps does not result in higher fidelity; instead, it reaches a plateau and oscillates around a fixed value. 
  }

 \label{fig:fid_depol}
\end{figure}

In this case, we applied gradient descent for 2000 iterations with a learning rate \(\eta = 0.3\). Further optimization does not yield an improvement, since the cost function converges. In fact, after approximately 300 iterations, 
the cost function begins to oscillate around \(4.1107\cdot10^{-5}\), preventing any further performance gains. Nevertheless, we continued the optimization up to 2000 iterations for illustrative purposes. On the other hand, in this case, the fidelity stabilizes around \(0.99586\). The training set in this setting consisted of the five MUBs for 2 qubits, represented in Eq.~\eqref{eq:MUBs}.
The evolution of the cost function is shown in Fig.~\ref{fig:cost_depol}. Furthermore, the evolution of the average fidelity is presented in Fig.~\ref{fig:fid_depol}.

After the optimization process, we compared the average fidelity over the MUB states obtained by the variational model with that of the QFT, both under the influence of depolarizing noise:

\begin{itemize}
    \item \textbf{Noisy QFT:} Average fidelity over the MUB states: \(0.99586 \pm 0.00007\).
    \item \textbf{Trained noisy variational circuit:} Average fidelity over the MUB states: \(0.99360 \pm 0.00009\).
\end{itemize}

After training, the noisy variational circuit achieves a (very slightly) lower average fidelity than the noisy QFT. This outcome indicates that, under these conditions, the variational circuit does not offer any advantage over directly implementing the theoretical QFT.

Additionally, we conducted the same numerical experiment for different values of the depolarizing parameter \( \epsilon \), spanning different orders of magnitude: 
\[ \epsilon = [10^{-8}, 10^{-7}, 10^{-6}, 10^{-5}, 10^{-4}, 10^{-3}, 10^{-2}, 10^{-1}]. \]

\begin{figure}[]
    \centering
    \text{Comparison of average fidelities} \\ \vspace{0.2cm}
        \includegraphics[scale=0.34]{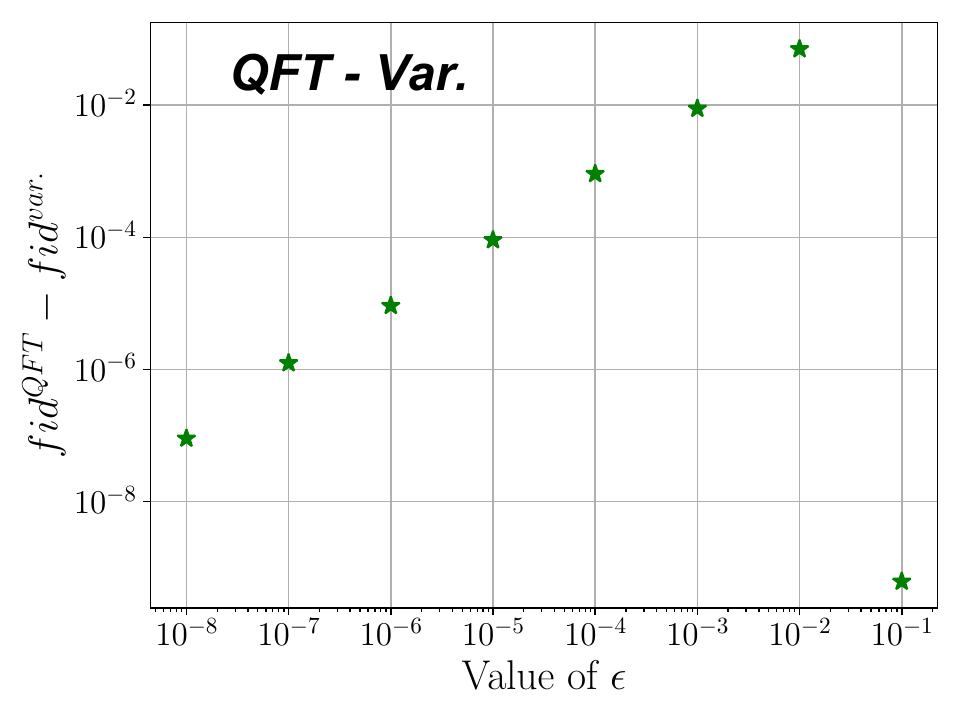}
  \caption{\small Difference in average fidelity, using a logarithmic scale, between the theoretical QFT and the variational circuit in the presence of depolarizing noise for different values of the depolarization parameter. For all these values, the variational circuit was unable to achieve a higher average fidelity than the original circuit.}

 \label{fig:dif_depol}
\end{figure}

For all these values, the variational circuit was unable to achieve a higher average fidelity than the theoretical QFT in the presence of depolarizing noise with the corresponding value of \( \epsilon \). The difference \(\text{fid}^{\text{QFT}} - \text{fid}^{\text{var.}} \) is shown in Fig.~\ref{fig:dif_depol}, where a positive value indicates that the variational circuit achieves a lower fidelity. It is worth noting that the point at \(\epsilon=0.1\) deviates from the trend, showing a smaller fidelity difference. This occurs because, at such a high depolarizing parameter, the states from both the variational circuit and the QFT converge to the maximally mixed two-qubit state, expressed as \(\tfrac{1}{4}\mathbb{I}_{4\times4}\), where \(\mathbb{I}_{4\times4}\) denotes the \(4\times4\) identity matrix, as confirmed by our simulations.

For current devices, if we identify \( \epsilon \) with a single-qubit gate error, the value of \( \epsilon \) typically lies between \( 10^{-4} \) and \( 10^{-3} \).

As in the noiseless case, we also computed the average fidelity of the trained variational circuit using 1000 randomly chosen superposition input states. The resulting average fidelities are:

\begin{itemize}
    \item \textbf{Noisy QFT:} Average fidelity over superposition states: \(0.99591 \pm 0.00001\).
    \item \textbf{Trained noisy variational circuit:} Average fidelity over superposition states: \(0.99367 \pm 0.00004\).
\end{itemize}

\begin{figure}[]
    \centering
    \text{Histogram Fidelities -- Depolarizing} \\ \vspace{0.2cm}
        \includegraphics[scale=0.4]{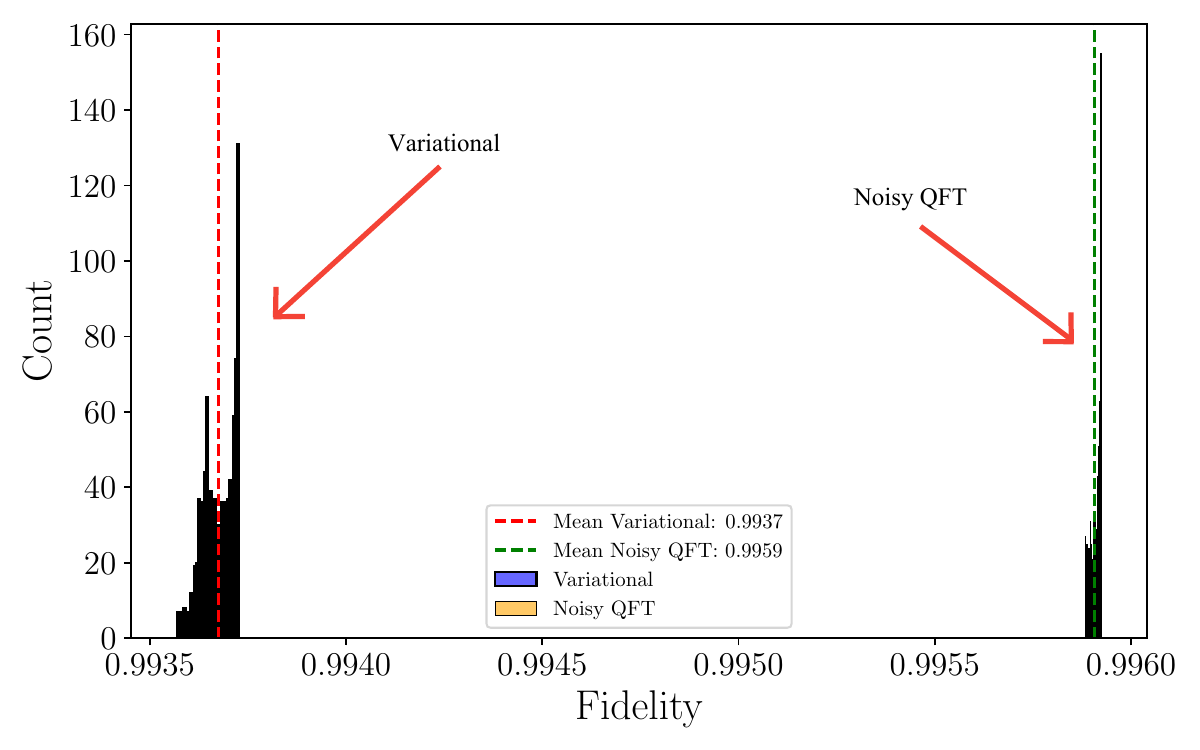}
  \caption{\small Distribution of fidelity values for 1000 random initial superposition states for the variational circuit (blue) using the optimal parameters obtained from the optimization and for the noisy QFT (orange). The dashed red line represents the average fidelity for the variational circuit, and the dashed green line is the average fidelity for the noisy QFT.}

 \label{fig:distr_depol}
\end{figure}

Additionally, we plot the distribution of fidelities for both the variational circuit and the noisy QFT in Fig.~\ref{fig:distr_depol}. Focusing solely on the variational circuit histogram, it is remarkable how the general behavior of random states can be explored by restricting only to MUBs. Nevertheless, it is evident that the noisy QFT still achieves higher fidelity across all random superposition states and exhibits a narrower distribution. Thus, the variational circuit does not offer an advantage over directly implementing the theoretical QFT in this scenario. 

\subsection{Depolarizing noise with single-qubit thermal relaxation}\label{dp_thermal_section}

\begin{figure}[]
    \centering
    \text{Cost -- Depolarizing \& Thermal Relaxation} \\ \vspace{0.2cm}
        \includegraphics[scale=0.34]{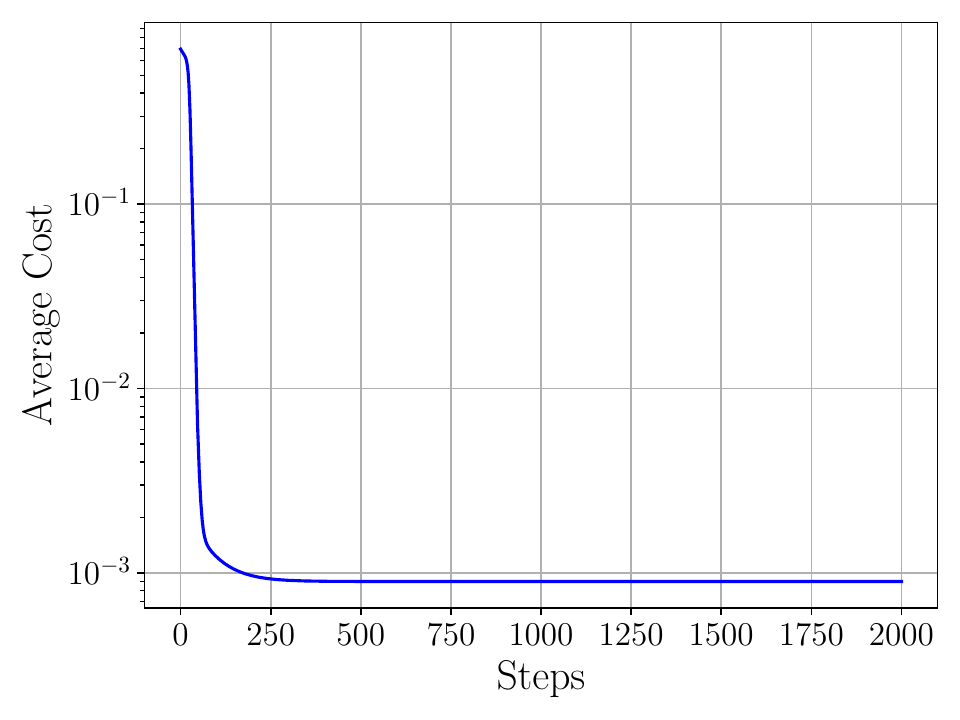}
  \caption{\small  Evolution of the cost function in the presence of depolarizing and thermal relaxation noise during optimization over 2000 steps. In this setting, training for additional steps does not lead to higher fidelity; instead, it reaches a plateau and oscillates around a fixed value.}

 \label{fig:cost_th}
\end{figure}

\begin{figure}[]
    \centering
    \text{Average Fidelity -- Depolarizing \& Thermal Relaxation} \\ \vspace{0.2cm}
        \includegraphics[scale=0.34]{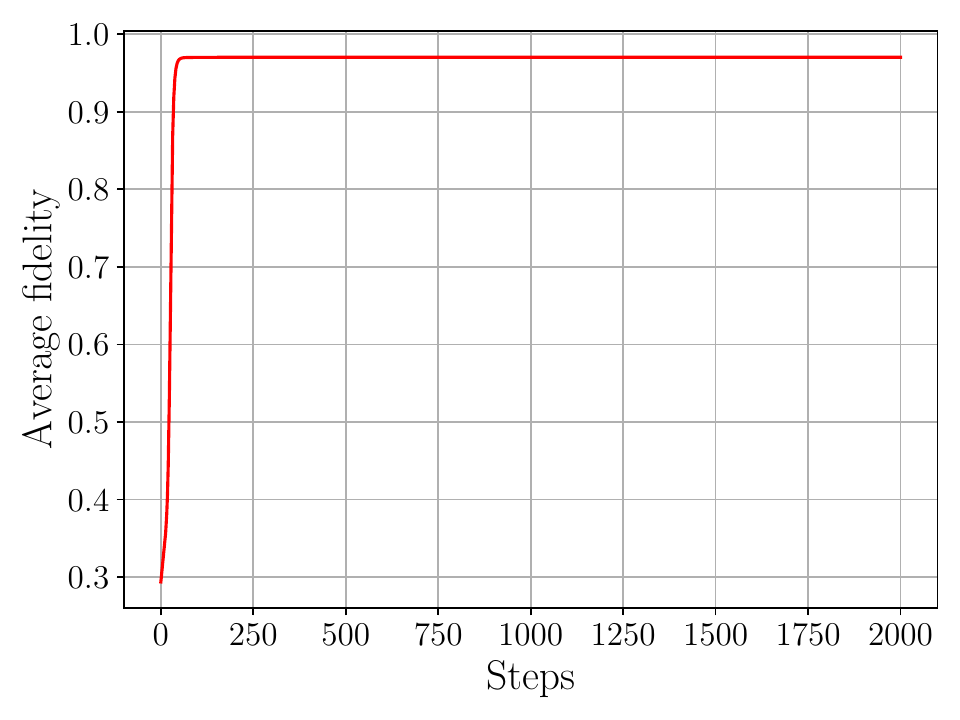}
  \caption{\small Evolution of the average fidelity during optimization in the presence of depolarizing and thermal relaxation noise during optimization over 2000 steps. In this setting, training for additional steps does not result in higher fidelity; instead, it reaches a plateau and oscillates around a fixed value.}

 \label{fig:fid_th}
\end{figure}

In the subsequent noise scenario, we introduced depolarizing noise combined with single-qubit thermal relaxation, using physical parameters extracted from the calibration data of the IBM quantum device \texttt{ibm\_brisbane} (see the table in the Appendix \ref{Calibration Data}). As in previous experiments, we employed gradient descent with a learning rate of \( \eta = 0.3 \) over 2000 iterations. Similar to the previous noise scenario, the optimization converges after approximately 300 iterations, and the cost function oscillates around \(7.88\cdot10^{-4}\), showing no further improvement. In its turn, the fidelity stabilizes around \(0.9721\). The evolution of the cost function throughout the optimization process is shown in Fig.~\ref{fig:cost_th}, while Fig.~\ref{fig:fid_th} presents the corresponding evolution of the average fidelity.

After completing the optimization, we evaluated the performance of the variational model by comparing its average fidelity to that of the noisy QFT. Both fidelities were computed with respect to the ideal QFT using the MUB states:

\begin{itemize}
    \item \textbf{Noisy QFT:} Average fidelity over the MUB states: \( 0.980 \pm 0.002 \).
    \item \textbf{Trained noisy variational circuit:} Average fidelity over the MUB states: \( 0.970 \pm 0.002 \).
\end{itemize}

In this noise scenario, we further evaluate the performance by calculating the average fidelity across 1000 randomly generated superposition states for both the optimized variational circuit and the QFT under noise. The corresponding average fidelities are:

\begin{itemize}
    \item \textbf{Noisy QFT:} Average fidelity over superposition states: \(0.981 \pm 0.002\).
    \item \textbf{Trained noisy variational circuit:} Average fidelity over superposition states: \(0.971 \pm 0.002\).
\end{itemize}

\begin{figure}[]
    \centering
    \text{Histogram Fidelities -- Depolarizing \& Thermal Relaxation} \\ \vspace{0.2cm}
        \includegraphics[scale=0.4]{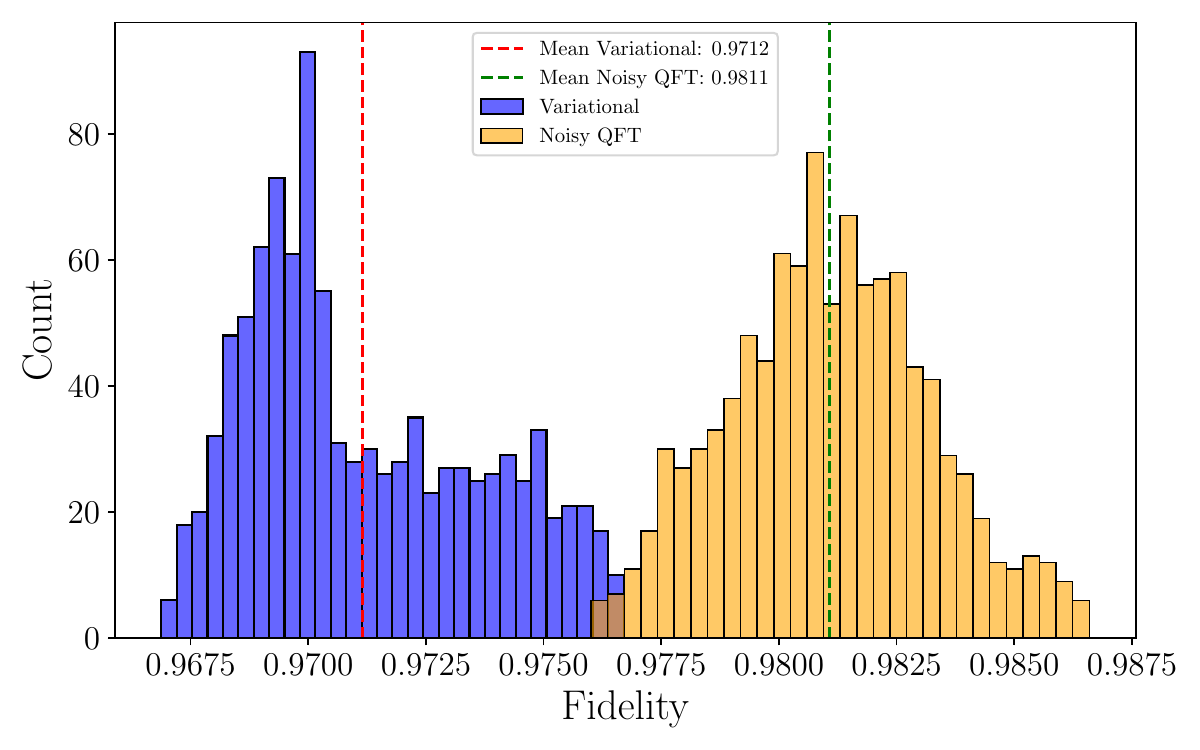}
  \caption{\small Distribution of fidelity values for 1000 random initial superposition states for the variational circuit (blue) using the optimal parameters obtained from the optimization and for the noisy QFT (orange). The dashed red line represents the average fidelity for the variational circuit, and the dashed green line is the average fidelity for the noisy QFT.}

 \label{fig:distr_th}
\end{figure}

The fidelity distributions for the various random superposition states are shown in Fig.~\ref{fig:distr_th}. It is evident that, in most cases, the noisy variational circuit underperforms compared to the noisy QFT. Therefore, in this scenario, employing the variational circuit offers no clear advantage over directly executing the theoretical QFT.

\subsection{Depolarizing noise and Crosstalk}\label{DP_CT_Results}

\begin{figure}[]
    \centering
    \text{Cost -- Depolarizing \& Crosstalk} \\ \vspace{0.2cm}
        \includegraphics[scale=0.34]{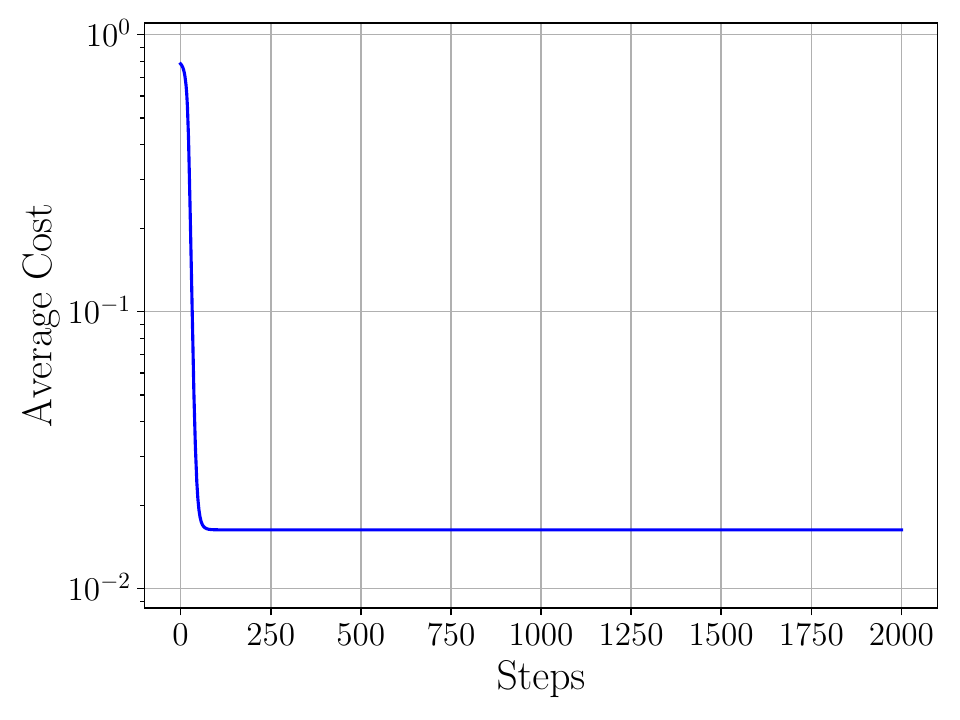}
  \caption{\small  Evolution of the cost function in the presence of depolarizing noise and crosstalk, presented on a logarithmic scale, during optimization over 2000 steps. In this setting, training for additional steps does not achieve a higher fidelity; instead, it reaches a plateau and starts oscillating around a given value.}

 \label{fig:cost_ct}
\end{figure}

\begin{figure}[]
    \centering
    \text{Average Fidelity -- Depolarizing \& Crosstalk} \\ \vspace{0.2cm}
        \includegraphics[scale=0.34]{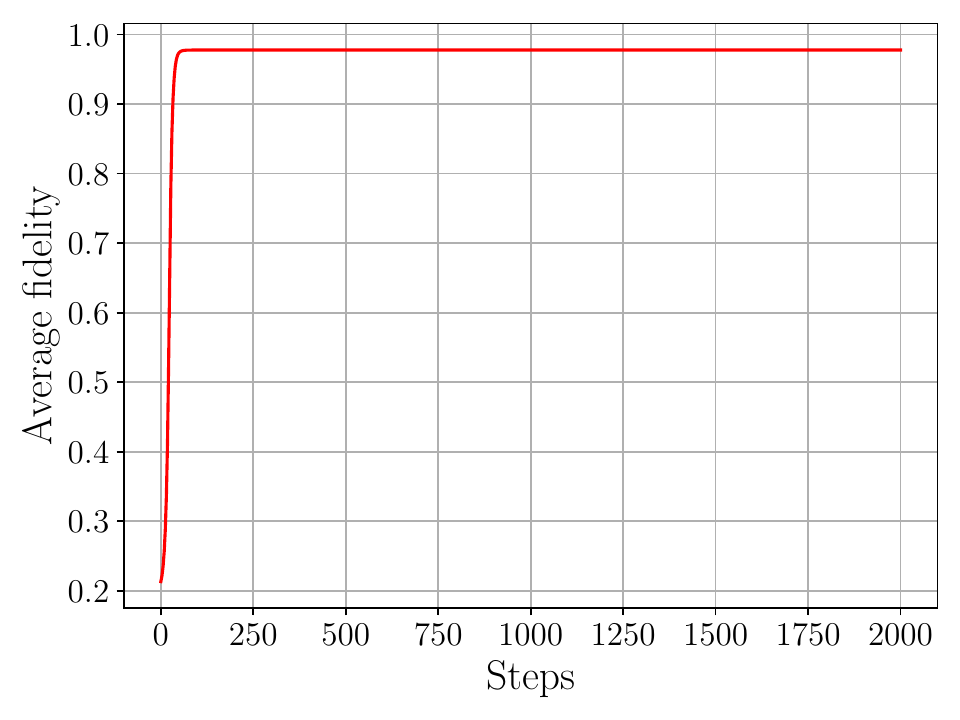}
  \caption{\small Evolution of the average fidelity during optimization in the presence of depolarizing and crosstalk noise over 2000 steps. In this setting, training for additional steps does not result in higher fidelity; instead, it reaches a plateau and oscillates around a fixed value.}

 \label{fig:fid_ct}
\end{figure}

In the third noise scenario, we introduced depolarizing noise combined with crosstalk noise between qubits. As in the previous numerical experiments, we employed gradient descent with a learning rate of \(\eta = 0.3\) over 2000 iterations. In this case, the cost function converges around \(1.96\cdot10^{-2}\), and the fidelity around \(0.9746\), showing no further improvement. The evolution of the cost function during optimization is depicted in Fig.~\ref{fig:cost_ct}, while Fig.~\ref{fig:fid_ct} shows the evolution of the average fidelity.

Upon completing the optimization, we assessed the performance of the variational model by comparing its average fidelity to that of the noisy QFT, both evaluated against the ideal QFT using the MUB states:

\begin{itemize}
    \item \textbf{Noisy QFT:} Average fidelity over the MUB states: \( 0.878 \pm 0.075 \).
    \item \textbf{Trained noisy variational circuit:} Average fidelity over the MUB states: \( 0.978 \pm 0.006 \).
\end{itemize}

The variational circuit outperforms the noisy QFT across the MUB states, exhibiting both higher average fidelity and reduced variance. In terms of \emph{average infidelity}—defined as \(1 - F_{avg}\), where \(F_{avg}\) denotes the average fidelity—this corresponds to a reduction from \(0.122\) for the noisy QFT to \(0.022\), representing approximately a fivefold suppression. These results indicate that the variational circuit is more resilient to the considered noise model and can partially mitigate its detrimental effects.

In this scenario, the variational circuit demonstrates a clear advantage over executing the theoretical QFT.
Similar to the previous case, we conducted experiments for different values of the depolarizing parameter \( \epsilon \), using the same range of values as before.

\begin{figure}[]
    \centering
    \text{Comparison of average fidelities} \\ \vspace{0.2cm}
        \includegraphics[scale=0.34]{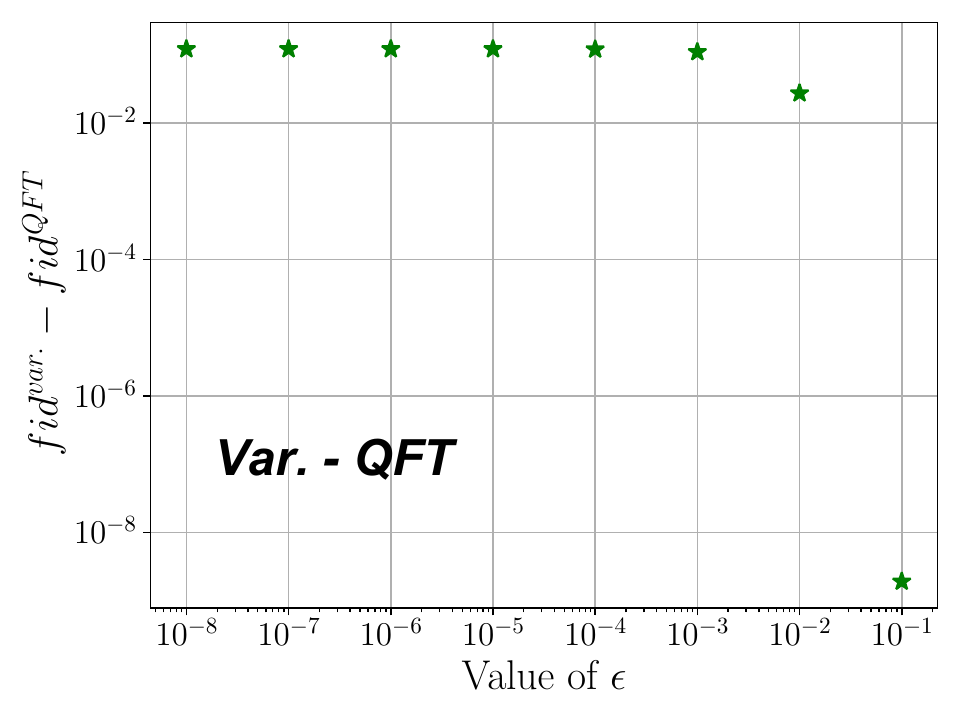}
  \caption{\small Difference in average fidelity between the theoretical QFT and the variational circuit under depolarizing and crosstalk noise for various values of the depolarization parameter. For most tested values, the variational circuit outperforms the theoretical QFT, achieving a higher average fidelity. The plot also shows that the fidelity difference is largely independent of \(\epsilon\), indicating that crosstalk noise dominates.}
\label{fig:dif_ct}
\end{figure}

For all tested values, the variational circuit achieved a higher average fidelity than the theoretical QFT. The difference \( \text{fid}^{\text{var.}} - \text{fid}^{\text{QFT}} \) is shown in Fig.~\ref{fig:dif_ct}, where a positive value indicates superior performance by the variational circuit. Note that in this case the difference is computed by subtracting the fidelity of the QFT from that of the variational circuit—opposite to the convention used in the depolarizing noise scenario—allowing us to display the data on a logarithmic scale. 

Additionally, the plot shows that the fidelity difference is almost independent of the depolarizing parameter \(\epsilon\), indicating that crosstalk noise dominates for most values. As in the previous noise scenario, the point corresponding to \(\epsilon=0.1\) deviates from the trend. This occurs because, at such a high depolarizing parameter, the states produced by both the variational circuit and the QFT converge to the maximally mixed two-qubit state.

In this setting, we also compute the average fidelity over 1000 random superposition states for both the trained variational circuit and the QFT under the presence of noise. The resulting average fidelities are:

\begin{itemize}
    \item \textbf{Noisy QFT:} Average fidelity over superposition states: \(0.885 \pm 0.062\).
    \item \textbf{Trained noisy variational circuit:} Average fidelity over superposition states: \(0.977 \pm 0.004\).
\end{itemize}

\begin{figure}[]
    \centering
    \text{Histogram Fidelities -- Depolarizing \& Crosstalk} \\ \vspace{0.2cm}
    \begin{minipage}{0.48\textwidth}
        \centering
        \includegraphics[scale=0.40]{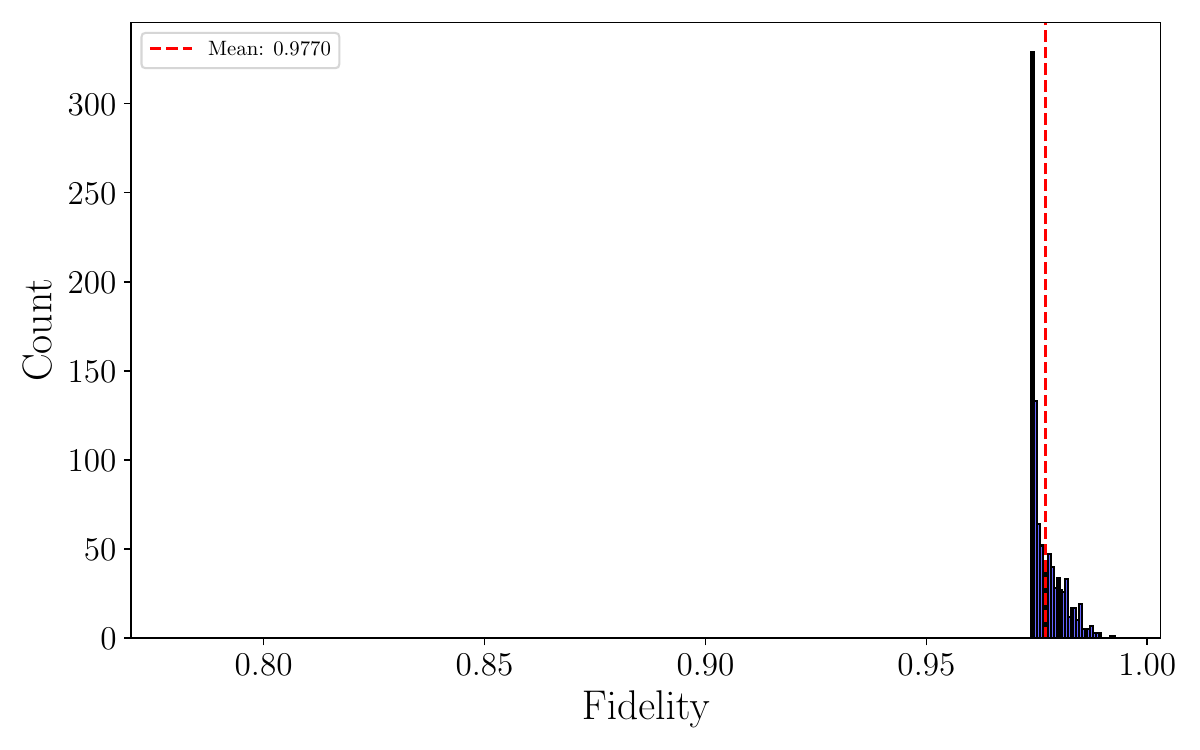}
        \end{minipage}
    \hfill
    \begin{minipage}{0.48\textwidth}
        \centering
        \includegraphics[scale=0.4]{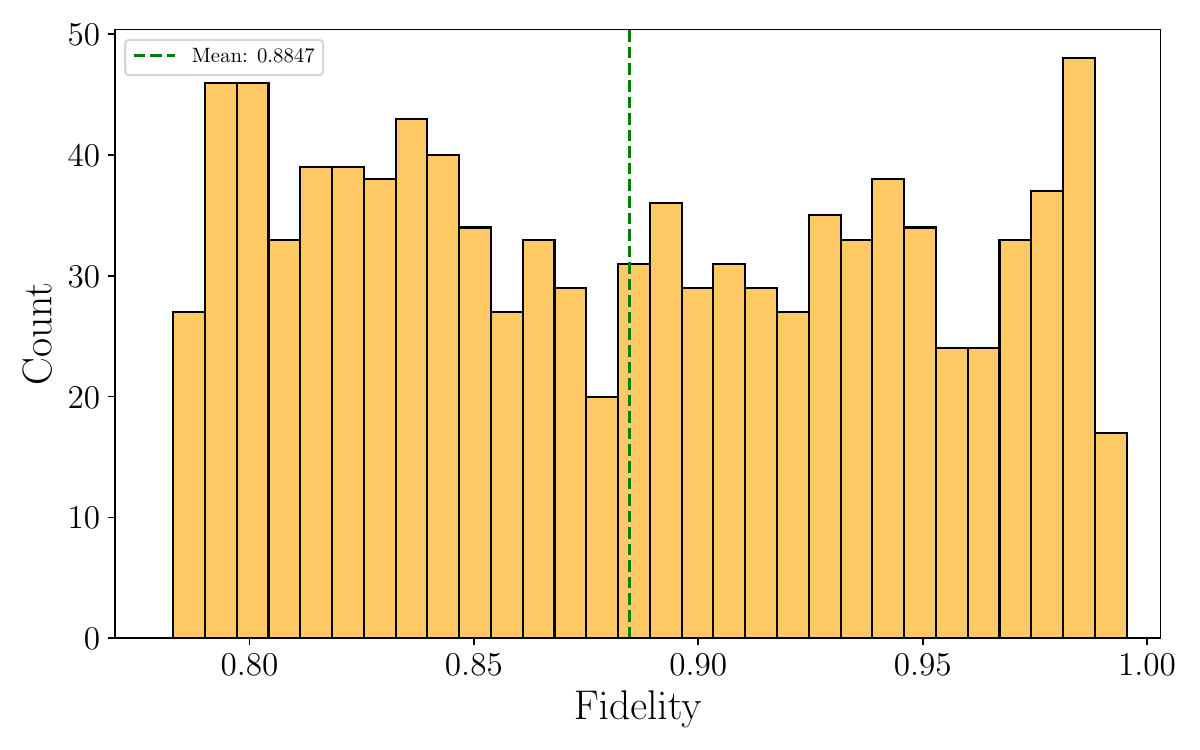}
        \caption{\small Distribution of fidelity values for 1000 random initial superposition states. \textbf{Upper panel:} Fidelity distribution for the variational circuit, with the dashed red line indicating the average fidelity. \textbf{Lower panel:} Fidelity distribution for the noisy QFT, with the dashed green line representing the average fidelity.}
        \label{fig:distr_ct}
    \end{minipage}
\end{figure}

The distributions of fidelities for the different random superposition states are shown in Fig.~\ref{fig:distr_ct}. For clarity, the distributions are plotted separately while maintaining a consistent x-axis range. The variational circuit outperforms the noisy QFT for the vast majority of random superposition states, as reflected in the higher average fidelity and narrower distribution. It achieves a fivefold suppression of average infidelity, from \(0.115\) for the noisy QFT to \(0.023\). The narrower distribution further indicates enhanced robustness of the variational circuit to the considered noise model.

\subsection{Depolarizing noise with single-qubit thermal relaxation and crosstalk between qubits}

In the final noise scenario, we performed simulations incorporating all relevant noise sources: depolarizing noise with single-qubit thermal relaxation, as well as crosstalk between qubits.

\begin{figure}[]
    \centering
    \text{Cost -- All Noise Sources} \\ \vspace{0.2cm}
        \includegraphics[scale=0.34]{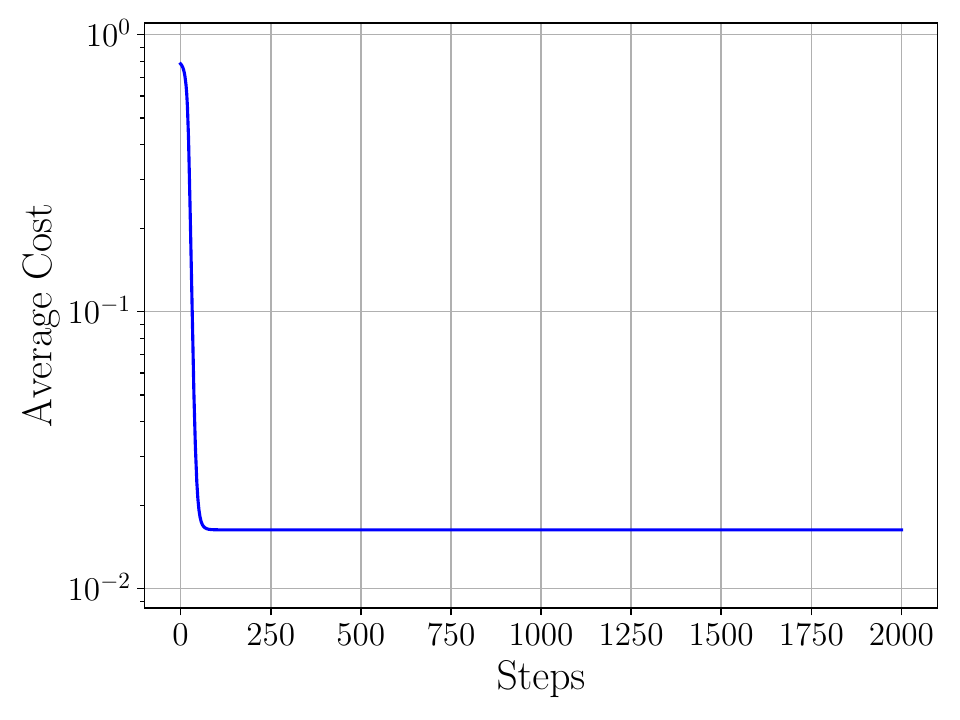}
  \caption{\small  Evolution of the cost function in the presence of depolarizing, crosstalk and thermal relaxation noise during optimization over 2000 steps. In this setting, training for additional steps does not lead to higher fidelity; instead, it reaches a plateau and oscillates around a fixed value.}

 \label{fig:cost_all}
\end{figure}

\begin{figure}[]
    \centering
    \text{Average Fidelity -- All Noise Sources} \\ \vspace{0.2cm}
        \includegraphics[scale=0.34]{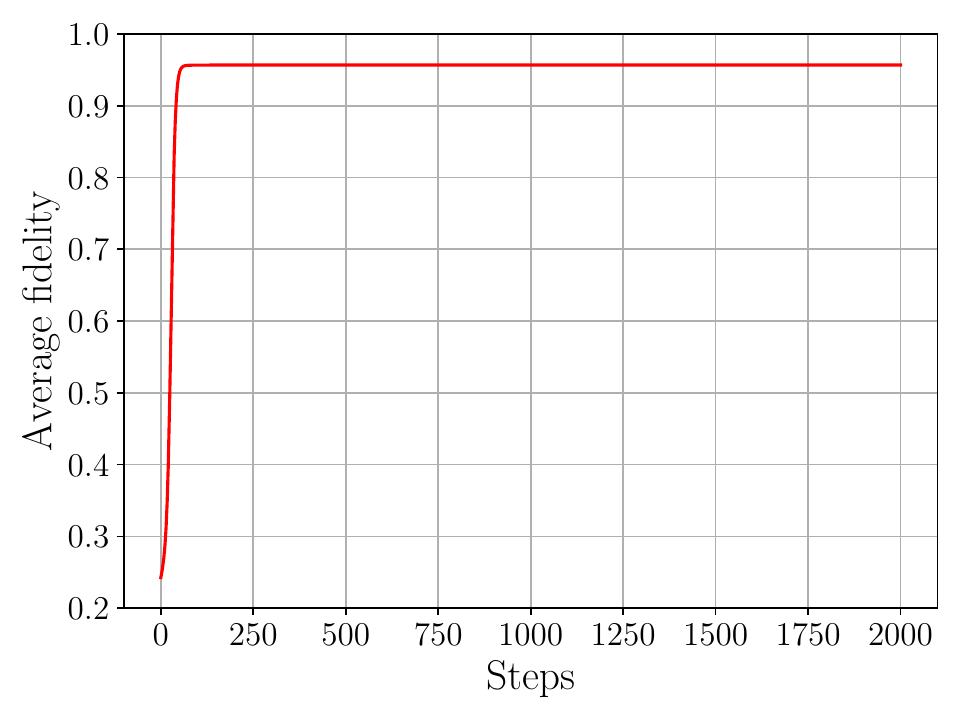}
  \caption{\small Evolution of the average fidelity during optimization in the presence of depolarizing, crosstalk and thermal relaxation noise during optimization over 2000 steps. In this setting, training for additional steps does not result in higher fidelity; instead, it reaches a plateau and oscillates around a fixed value.}

 \label{fig:fid_all}
\end{figure}

The noise parameters used in this setting are the same as those from the Sec. \ref{dp_thermal_section}, derived from real calibration data of the IBM quantum device \texttt{ibm\_brisbane} (see the table in Appendix~\ref{Calibration Data}). As in earlier experiments, gradient descent was applied with a learning rate of \(\eta = 0.3\) for 2000 iterations. In this case, the convergence criterion with \(\delta = 10^{-9}\) would be satisfied after approximately 950 iterations, after which the optimization becomes trapped and the cost function oscillates around \(2.14\cdot10^{-2}\) without further improvement. The optimization was nonetheless extended to 2000 iterations to demonstrate this behavior. Consistent with previous cases, a zoomed view of the fidelity is not shown, as it would mainly reveal numerical fluctuations around \(0.9536\). The evolution of the cost function throughout the optimization process is shown in Fig.~\ref{fig:cost_all}, while Fig.~\ref{fig:fid_all} presents the corresponding evolution of the average fidelity.

After the optimization process, we assessed the performance of the variational model by comparing its average fidelity to that of the noisy QFT, with both fidelities evaluated against the ideal QFT using the MUB states:

\begin{itemize}
    \item \textbf{Noisy QFT:} Average fidelity over the MUB states: \( 0.866 \pm 0.074 \).
    \item \textbf{Trained noisy variational circuit:} Average fidelity over the MUB states: \( 0.957 \pm 0.006 \).
\end{itemize}

The variational circuit outperforms the noisy QFT by exhibiting both higher average fidelity and reduced variance. In terms of infidelity, this corresponds to a reduction from \(0.134\) for the noisy QFT to \(0.043\), representing approximately a threefold suppression of error. These results indicate that the variational approach is more robust to the considered noise model and can partially mitigate its detrimental effects.

As observed in the depolarizing and crosstalk noise scenario, the variational circuit offers a clear advantage over directly executing the theoretical QFT circuit.

We further assess performance by computing the average fidelity over 1000 randomly generated superposition states for both the optimized variational circuit and the noisy QFT. The resulting average fidelities are:

\begin{itemize}
    \item \textbf{Noisy QFT:} Average fidelity over superposition states: \(0.868 \pm 0.063\).
    \item \textbf{Trained noisy variational circuit:} Average fidelity over superposition states: \(0.956 \pm 0.004\).
\end{itemize}

\begin{figure}[]
    \centering
    \text{Histogram Fidelities -- All Noise Sources} \\ \vspace{0.2cm}
    \begin{minipage}{0.48\textwidth}
        \centering
        \includegraphics[scale=0.40]{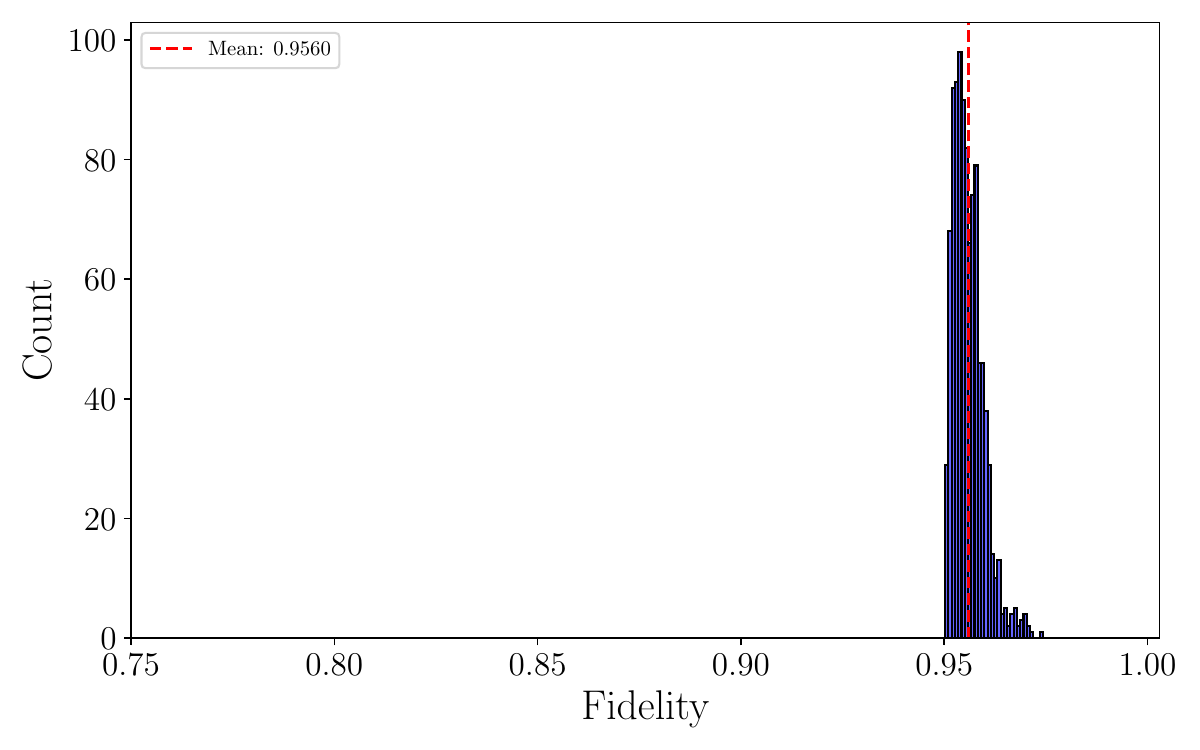}
        \end{minipage}
    \hfill
    \begin{minipage}{0.48\textwidth}
        \centering
        \includegraphics[scale=0.4]{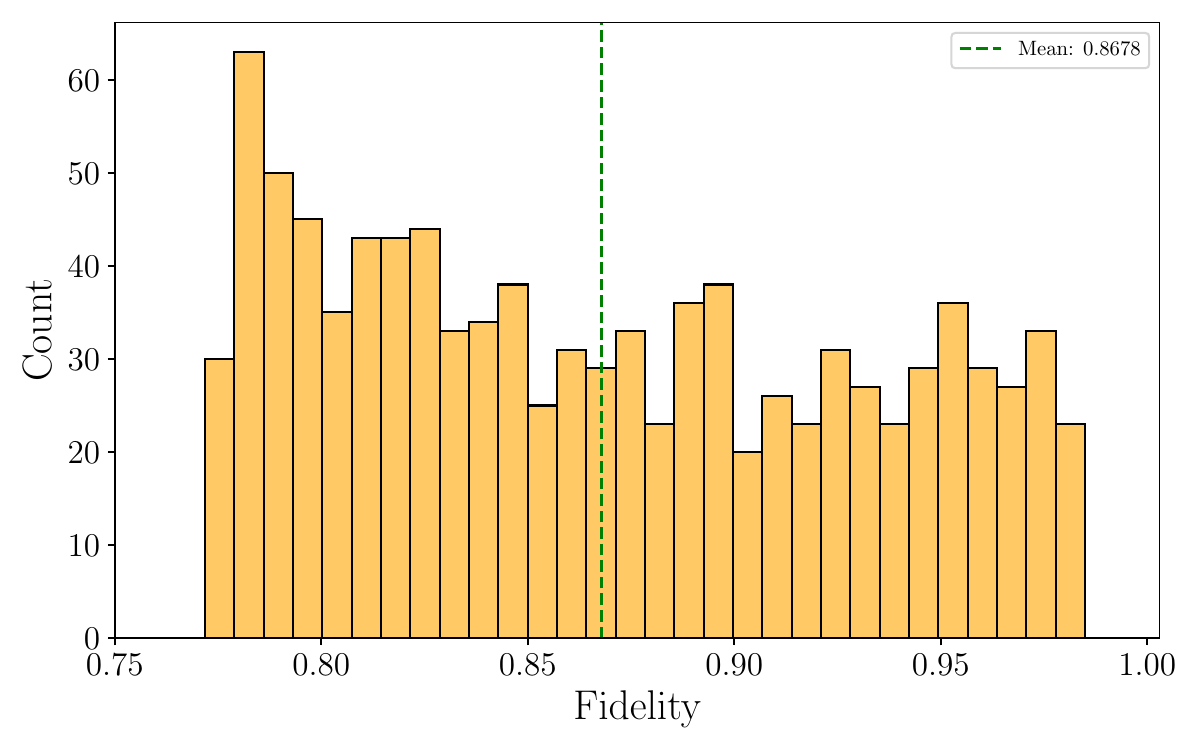}
        \caption{\small Distribution of fidelity values for 1000 random initial superposition states. \textbf{Upper panel:} Fidelity distribution for the variational circuit, with the dashed red line indicating the average fidelity. \textbf{Lower panel:} Fidelity distribution for the noisy QFT, with the dashed green line representing the average fidelity.}
        \label{fig:distr_all}
    \end{minipage}
\end{figure}

Figure~\ref{fig:distr_all} shows the fidelity distributions for 1000 random superposition states. For higher clarity, the distributions are displayed separately while keeping the same x-axis range. The variational circuit outperforms the noisy QFT in the majority of cases. In terms of infidelity, this corresponds to a reduction from \(0.132\) for the noisy QFT to \(0.044\), representing approximately a threefold suppression of infidelity. Additionally, the narrower spread of the variational circuit’s distribution suggests enhanced robustness against this specific type of noise.

\section{Conclusions and outlook}\label{Conclusions}

This work evaluated the ability of a variational quantum circuit to reproduce the Quantum Fourier Transform (QFT) in the presence of noise with higher fidelity than the theoretical quantum circuit.

The results show that the variational quantum circuit achieves higher fidelity than the QFT in noise scenarios that include coherent noise. These scenarios also involve non-coherent noise. Specifically, for the initial MUB states, the variational circuit reduces the infidelity by approximately a factor of five in the depolarizing and crosstalk noise scenario, and by about a factor of three when all noise sources are present. For random superposition states, the infidelity is reduced by roughly a factor of five and a factor of three in the respective noise scenarios.

These findings highlight the potential of this protocol as an error-mitigating tool, particularly for small-to-medium-scale quantum systems in scenarios where coherent noise plays a significant role. The approach provides an effective strategy to mitigate the impact of such noise and to enhance fidelity. Furthermore, it can be tailored to the specific noise profile of a given quantum device, offering a flexible and practical solution for improving the performance of quantum algorithms under noisy conditions.

Additionally, as a by-product of our investigation, we discovered the effectiveness of using the Mutually Unbiased Basis (MUB) as a training set for exploring the quantum state space during optimization. The use of MUBs allowed us to achieve not only higher average fidelities but also reduced variance, helping to avoid certain local minima. We believe this novel approach could be extended to other optimization tasks in Quantum Optimization and Quantum Machine Learning, serving as a valuable tool for efficiently exploring the parameterized quantum state space.


In this work, the numerical experiments were exclusively based on classical simulations of a quantum system to test the protocol. Future work should investigate the performance of the method under more sophisticated noise models. In particular, the approach could be extended to circuits with a larger number of qubits by exploiting the recursive construction of the QFT circuit discussed in Sec.~\ref{Theory}. Moreover, the protocol could be benchmarked within established quantum algorithms, such as by integrating the optimized variational QFT as a subroutine in the Quantum Phase Estimation algorithm ~\cite{kitaev1995quantummeasurementsabelianstabilizer}, to evaluate its effectiveness in more realistic computational settings. 

\section*{Acknowledgements}
This work was supported by the project PID2023-152724NA-I00, with funding from MCIU/AEI/10.13039/501100011033 and FSE+,
the Severo Ochoa Grant CEX2023-001292-S, Generalitat Valenciana grant CIPROM/2022/66, the Ministry of Economic
Affairs and Digital Transformation of the Spanish Government through
the QUANTUM ENIA project call - QUANTUM SPAIN project, and by the
European Union through the Recovery, Transformation and Resilience
Plan - NextGenerationEU within the framework of the Digital Spain
2026 Agenda, and by the CSIC Interdisciplinary Thematic Platform (PTI+)
on Quantum Technologies (PTI-QTEP+). This project has also received
funding from the European Union’s Horizon 2020 research and innovation
program under grant agreement CaLIGOLA MSCA-2021-SE-01-101086123. RGL is funded by grant CIACIF/2021/136 from Generalitat Valenciana. BZ
acknowledges the support from Generalitat Valenciana through the “GenT program”, ref.:
CIDEGENT/2020/055. The work of A.B. is supported through the FPI grant PRE2020-095867 funded by MCIN/AEI/10.13039/501100011033

\

\bibliographystyle{unsrt}
\bibliography{references}

\onecolumngrid
\appendix
\section{Calibration data from the IBM device \texttt{ibm\_brisbane}}\label{Calibration Data}

In this Appendix, we provide the calibration data from the IBM device \texttt{ibm\_brisbane}, which were used as parameters for the noise model, specifically for the inclusion of depolarizing noise and thermal relaxation effects during circuit execution. The calibration data is shown in Table \ref{tab:qubit23_data}.

\begin{table}[h!]
\centering
\resizebox{\textwidth}{!}{%
\begin{tabular}{@{}llllllllllll@{}}
\toprule
Qubit & T1 (\textmu s) & T2 (\textmu s) & Frequency (GHz) & Temperature (mK) & Single-Qubit Gate Time (ns) & ID error  & RZ error & SX error  & X error   & ECR error         & 2-qubit Gate time (ns)    \\ \midrule
2     & 316.85         & 311.90         & 4.61           & 15           & 60                   & 0.0002045 & 0        & 0.0002045 & 0.0002045 & --                & --                \\
3     & 315.50         & 409.37         & 4.86          & 15             & 60                  & 0.0005032 & 0        & 0.0005032 & 0.0005032 & ecr3\_2:0.008892    & ecr3\_2:660    \\ 
\bottomrule
\end{tabular}%
}
\caption{Device calibration parameters for qubits 2 and 3 from \texttt{ibm\_brisbane}.}
\label{tab:qubit23_data}
\end{table}

\end{document}